\documentclass[aps,pra]{revtex4}
\usepackage{amssymb}
\usepackage{graphicx}
\usepackage{dcolumn}
\usepackage{bm}

\begin{document}

\title{Interaction-induced interference for two independent Bose-Einstein
condensates}
\author{Hongwei Xiong$^{1,2,3}$\footnote {xionghongwei@wipm.ac.cn }, Shujuan Liu$^{1,2,3}$, Mingsheng Zhan$%
^{1,2} $}

\address{$^{1}$State Key Laboratory of Magnetic Resonance and
Atomic and Molecular Physics, Wuhan Institute of Physics and
Mathematics, Chinese Academy of Sciences, Wuhan 430071, P. R. China}
\address{$^{2}$Center for Cold Atom Physics, Chinese Academy of Sciences, Wuhan 430071, P.
R. China}
\address{$^{3}$Graduate School of the Chinese Academy of
Sciences, P. R. China}

\date{\today }

\begin{abstract}
After removing the double-well potential trapping two initially independent
Bose condensates, the density expectation value is calculated when both the
exchange symmetry of identical bosons and interatomic interaction are
considered. The density expectation value and evolution equations are
obtained based on both the first-quantization and second-quantization
methods. When the interatomic interaction is considered carefully, after the
overlapping of two initially independent condensates, it is shown that there
is a nonzero interference term in the density expectation value. It is found
that the calculated density expectation value with this model agrees with
the interference pattern observed in the experiment by Andrews \textit{et al
} \textit{(Science \textbf{275}, 637 (1997))}. The nonzero interference term
in the density expectation value physically arises from the exchange
symmetry of identical bosons and interatomic interaction which make two
initially independent condensates become coherent after the overlapping. For
two initially independent condensates, our researches show that there is an
interaction-induced coherence process.

\end{abstract}

\maketitle

\section{\protect\bigskip introduction}

The coherence property plays an essential role in the wave nature of
Bose-Einstein condensates (BECs), and it has been investigated intensively
\cite{Nature} after the experimental realizations of BECs in dilute gases
\cite{Anderson,Davis,Bradley}. For a single BEC which is a perfect quantum
fluid at zero temperature, the existence of a macroscopic wave function (or
order parameter) means that spatial coherence is an intriguing property of
the condensate. In the dilute Bose-condensed gases, the spatial coherence
can be investigated directly by interfering two Bose condensates. For two
coherently separated BECs, because the relative phase of two sub-condensates
is locked, it is not surprising that there is a clear interference pattern
when the two sub-condensates are allowed to overlap. In the celebrated
experiment by Andrews \textit{et al }\cite{Andrew}, however, high-contrast
fringes were observed even for two completely independent condensates at an
initial time. In this experiment, to prepare two initially independent
condensates, the dilute Bose gases were evaporatively cooled in a
double-well potential created by splitting a magnetic trap in half with a
far-off blue-detuned laser beam (See also \cite{ketterle}). In particular,
the height of the external potential due to the laser beam is much larger
than the chemical potential of two separated condensates, which means that
the tunneling current can be safely omitted. After switching off the
double-well potential, the two initially independent condensates overlapped
and high-contrast fringes were observed in Ref. \cite{Andrew}. This
experimental result shows clearly that there is a spatial coherence property
after the overlapping between two independent interacting condensates.

Although high-contrast fringes were observed for two initially independent
condensates, in many literature (see for example \cite%
{Leggett,Pethick,Stringari} and references therein), it is shown that there
is no interference term in the density expectation value for two initially
independent condensates. To solve the contradiction between this result and
the observed high-contrast fringes \cite{Andrew} for two initially
independent condensates, the observed high-contrast fringes were interpreted
with the aid of the high-order correlation function $P\left( \mathbf{r},%
\mathbf{r}^{\prime },t\right) $ (which is an oscillation function of $%
\mathbf{r}-\mathbf{r}^{\prime }$) and quantum measurement process in the
present popular viewpoint. Several theories have been proposed to interpret
the observed high-contrast fringes for two initially independent condensates
such as the stochastic simulations of the photon detection for atoms \cite%
{JAV}, the expansion of Fock state by the linear superposition of coherent
states \cite{CASTIN}, and the continuous measurement theory \cite{Zoller}.

In the present work, we calculate the density expectation value for two
initially independent condensates by including carefully the interatomic
interaction. Quite different from the simple derivation in other theories,
it is found that \textit{for the case of two initially independent
condensates, upon expansion, there is a nonzero interference term in the
density expectation value when the interatomic interaction and exchange
symmetry of identical bosons are both taken into account.} To exclude any
possible error in the cumbersome derivations, we calculate the density
expectation value and evolution equations by both the first-quantization and
second-quantization methods, and the same results are obtained. After
removing the double-well potential trapping the two initially independent
condensates, we give the theoretical result of the density expectation value
which agrees with the interference pattern observed in \cite{Andrew}.

To show the interaction-induced coherence process between two initially
independent Bose condensates when both the interatomic interaction and
exchange symmetry of identical bosons are considered carefully, the paper is
organized as follows. In Sec. II, we give the general expression of the
density expectation value for two initially independent condensates based on
the many-body wave function of the whole system. It is shown that there is a
nonzero interference term when the wave functions of two initially
independent condensates are no more orthogonal after their overlapping. In
Sec. III, we give the general expression of the overall energy of the whole
system, and the evolution equations are given by the action principle. In
Sec. IV, we give a brief proof why the wave functions of two initially
independent condensates become non-orthogonal after their overlapping when
the interatomic interaction is considered. In this section, an effective
order parameter is introduced to show further the interaction-induced
coherence formation process for two initially independent condensates. In
Sec. V, we give the numerical results for the evolution of the density
expectation value according to the experimental parameters in Ref. \cite%
{Andrew}. It is shown clearly that our theoretical results of the density
expectation value agree with the experimental results of the interference
patterns. In Sec. VI, the evolution of the density expectation value is
given for different coupling constants. It is shown that increasing the
coupling constant has the effect of enhancing the interference effect for
two independent condensates. To show further the physical mechanism of the
interaction-induced coherence process, in Sec. VII, we prove that the
theoretical results based on the second-quantization method is the same as
the results derived from the many-body wave function in the previous
sections. In Sec. VIII, the role of quantum fluctations and the
orthogonality of the many-body wave function are discussed. Finally, we give
a brief summary and discussion in the last section.

\section{the expression of the density expectation value for two initially
independent BECs}

First, we give a brief introduction to the scheme of observing the
interference effect of two separated condensates. As shown in Fig. 1, the
double-well potential can be created by superposing a far-off blue-detuned
laser beam (which generating a repulsive optical dipole force for atoms)
upon the magnetic trap. The combined double-well potential is%
\begin{equation}
V_{ext}=\frac{1}{2}m\left[ \omega _{x}^{2}\left( x-x_{0}\right) ^{2}+\omega
_{y}^{2}y^{2}+\omega _{z}^{2}z^{2}\right] +U_{0}e^{-\left( x-x_{0}\right)
^{2}/w_{x}^{2}-\left( y^{2}+z^{2}\right) /w_{\perp }^{2}},
\label{external potential}
\end{equation}%
where the first term is the trapping potential due to the magnetic trap,
while the second term represents the potential due to the laser beam.

\begin{figure}[tbp]
\includegraphics[width=0.8\linewidth,angle=270]{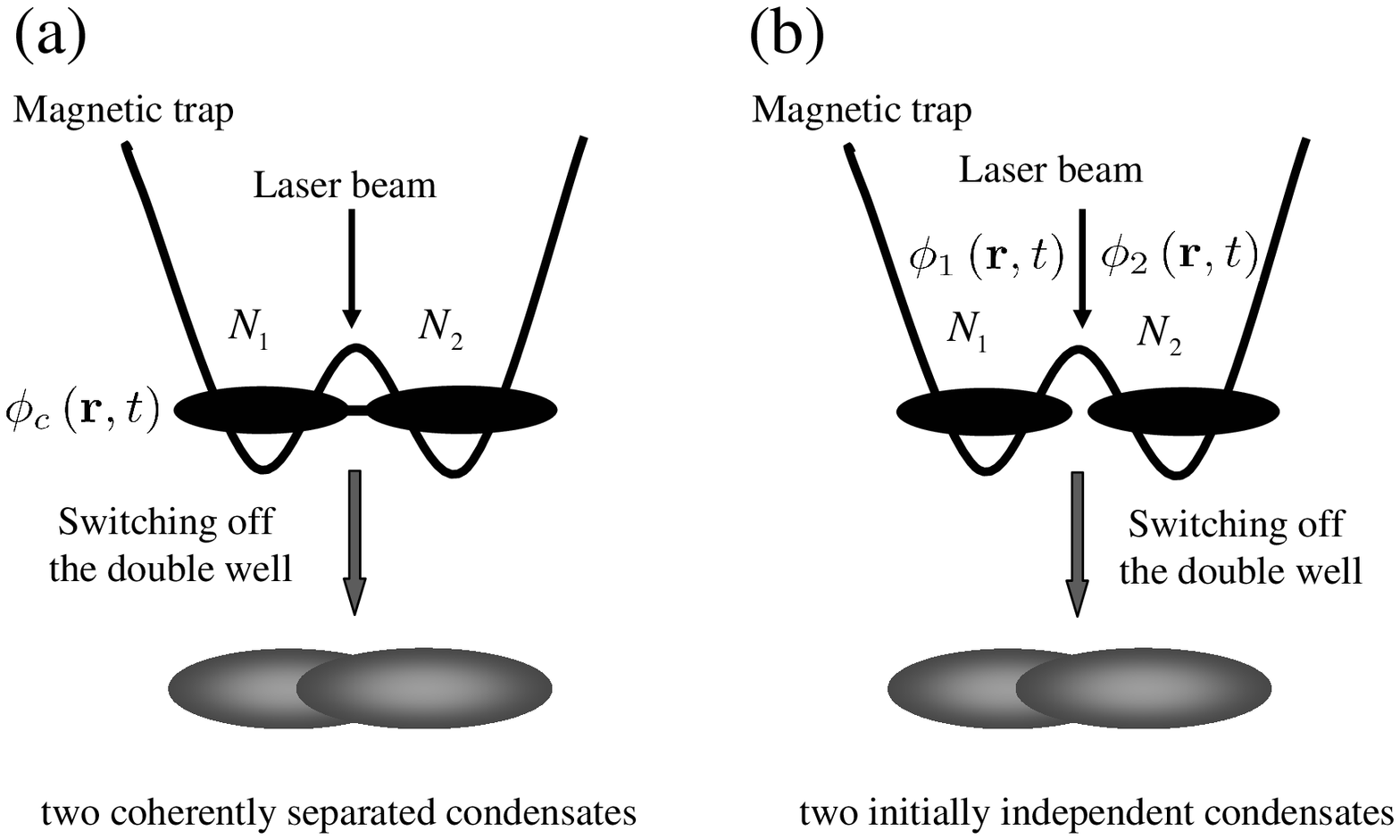}
\caption{The scheme of observing the interference effect of two separated
condensates. Fig.1(a) shows the case of two coherently separated
sub-condensates at an initial time, while Fig.1(b) shows two initially
independent condensates.}
\end{figure}

For the Bose gases confined in this double-well potential, there are two
quite different cases. (i) If the height $U_{0}$ of the external potential
due to the laser beam is smaller than the chemical potential of the system,
the two sub-condensates can be regarded to be coherently separated after the
evaporative cooling. In this situation, the two coherently-separated
sub-condensates can be regarded as a single condensate. The relative phase
of two sub-condensates is thus locked. (ii) If the height $U_{0}$ of the
laser beam is much larger than the chemical potential of the system so that
the tunneling effect can be omitted safely, the two condensates are
completely independent after the evaporative cooling.

For two coherently separated sub-condensates, every atom is described by the
following normalization wave function
\begin{equation}
\phi _{c}\left( \mathbf{r},t\right) =\left[ \sqrt{N_{1}}\phi _{c1}\left(
\mathbf{r},t\right) +\sqrt{N_{2}}\phi _{c2}\left( \mathbf{r},t\right) \right]
/\sqrt{N},
\end{equation}%
where $\phi _{c1}\left( \mathbf{r},t\right) $ and $\phi _{c2}\left( \mathbf{r%
},t\right) $ are the normalization wave functions accounting for the two
sub-condensates. $N=N_{1}+N_{2}$ is the total number of particles. Before
removing the double-well potential, the average particle numbers in each
condensate are $N_{1}$ and $N_{2}$, respectively. After removing the
double-well potential, the evolution of the wave function $\phi _{c}\left(
\mathbf{r},t\right) $ can be obtained based on the Gross-Pitaevskii (GP)
equation \cite{Gross,Pitae,RMP}. The density expectation value is then
\begin{eqnarray}
n_{c}\left( \mathbf{r},t\right) &=&N\left\vert \phi _{c}\left( \mathbf{r}%
,t\right) \right\vert ^{2}  \nonumber \\
&=&N_{1}\left\vert \phi _{c1}\left( \mathbf{r},t\right) \right\vert ^{2}+2%
\sqrt{N_{1}N_{2}}\times \mathrm{Re}\left[ \phi _{c1}^{\ast }\left( \mathbf{r}%
,t\right) \phi _{c2}\left( \mathbf{r},t\right) \right] +N_{2}\left\vert \phi
_{c2}\left( \mathbf{r},t\right) \right\vert ^{2}.  \label{coherent-density}
\end{eqnarray}%
The second term in the above equation accounts for the interference effect
when there is an overlapping between two sub-condensates upon expansion. For
$\phi _{c}\left( \mathbf{r},t\right) $ satisfying the GP equation, the
nonlinear effects in the interference pattern of two coherently separated
sub-condensates were investigated in Refs \cite{Rohrl,Liu}.

As shown in Fig. 1(b), if the intensity of the blue-detuned laser beam is
sufficiently high so that the tunneling effect can be omitted, the two
condensates can be regarded to be completely independent. In this situation,
the number of particles $N_{1}$ and $N_{2}$ in each of the two condensates
are fixed. We assume that there are $N_{1}$ particles described by the wave
function $\phi _{1}$ in the left well, while there are $N_{2}$ particles
described by the wave function $\phi _{2}$ in the right well. To investigate
clearly the role of the exchange symmetry of identical particles and
interatomic interaction, we calculate in the following the density
expectation value $n_{d}\left( \mathbf{r},t\right) $ directly from the
many-body wave function. In Sec. VII, we will also calculate the density
expectation value based on the second-quantization method, and the same
result is obtained.

We will prove in Sec. IV that in the presence of the interatomic
interaction, $\phi _{1}\left( \mathbf{r},t\right) $ and $\phi _{2}\left(
\mathbf{r},t\right) $ will become non-orthogonal after the overlapping
between two initially independent condensates. Thus we consider here the
following general case from the beginning%
\begin{equation}
\zeta \left( t\right) =\int \phi _{1}\left( \mathbf{r},t\right) \phi
_{2}^{\ast }\left( \mathbf{r},t\right) dV=\left\vert \zeta \left( t\right)
\right\vert e^{i\varphi _{c}}.  \label{xi}
\end{equation}%
After removing the double-well potential, the two initially independent
condensates will overlap, and thus one should consider the
indistinguishability of identical particles. When the exchange symmetry of
identical bosons is considered, the many-body wave function is%
\begin{eqnarray}
\Psi _{N_{1}N_{2}}\left( \mathbf{r}_{1},\mathbf{r}_{2},\cdots ,\mathbf{r}%
_{N},t\right) &=&A_{n}\sqrt{\frac{N_{1}!N_{2}!}{N!}}\sum\limits_{P}P\left[
\phi _{1}\left( \mathbf{r}_{1},t\right) \cdots \phi _{1}\left( \mathbf{r}%
_{N_{1}},t\right) \times \right.  \nonumber \\
&&\left. \phi _{2}\left( \mathbf{r}_{N_{1}+1},t\right) \cdots \phi
_{2}\left( \mathbf{r}_{N_{1}+N_{2}},t\right) \right] ,  \label{wave-function}
\end{eqnarray}%
where $P$ denotes the $N!/(N_{1}!N_{2}!)$ permutations for the particles in
different single-particle state $\phi _{1}$ or $\phi _{2}$. $A_{n}$ is a
normalization factor to assure $\int \left\vert \Psi _{N_{1}N_{2}}\left(
\mathbf{r}_{1},\mathbf{r}_{2},\cdots ,\mathbf{r}_{N},t\right) \right\vert
^{2}d\mathbf{r}_{1}d\mathbf{r}_{2}\cdots d\mathbf{r}_{N}=1$. $A_{n}$ is
determined by the following equation:
\begin{equation}
A_{n}\left[ \sum\limits_{i=0}^{\min \left( N_{1},N_{2}\right) }\frac{%
N_{1}!N_{2}!\left\vert \zeta \left( t\right) \right\vert ^{2i}}{i!i!\left(
N_{1}-i\right) !\left( N_{2}-i\right) !}\right] ^{1/2}=1.
\end{equation}%
In this paper, to give a concise expression for various coefficients such as
$A_{n}$, we have introduced the rule $0^{0}=1$. When the non-orthogonal
property between $\phi _{1}$ and $\phi _{2}$ is considered, one should note
that the normalization constant $A_{n}$ is relevant to the parameter $\zeta $%
. From the form of the many-body wave function
(\ref{wave-function}), the quantum depletion originating from
interparticle interaction is omitted. Thus, this form of quantum
state is valid when $a_{s}/\overline{l}<<1$ with $a_{s}$ and
$\overline{l}$ being respectively the scattering length and mean
distance between particles. The role of quantum depletion will be
discussed in Sec. VIII.

From the above many-body wave function, after straightforward derivations,
the exact expression of the density expectation value takes the following
form:
\begin{eqnarray}
n_{d}\left( \mathbf{r},t\right) &=&N\int \Psi _{N_{1}N_{2}}^{\ast }\left(
\mathbf{r},\mathbf{r}_{2},\cdots ,\mathbf{r}_{N},t\right) \Psi
_{N_{1}N_{2}}\left( \mathbf{r},\mathbf{r}_{2},\cdots ,\mathbf{r}%
_{N},t\right) d^{3}\mathbf{r}_{2}\cdots d^{3}\mathbf{r}_{N}  \nonumber \\
&=&a_{d}\left\vert \phi _{1}\left( \mathbf{r},t\right) \right\vert
^{2}+2b_{d}\times \mathrm{Re}\left[ e^{i\varphi _{c}}\phi _{1}^{\ast }\left(
\mathbf{r},t\right) \phi _{2}\left( \mathbf{r},t\right) \right]
+c_{d}\left\vert \phi _{2}\left( \mathbf{r},t\right) \right\vert ^{2},
\label{density}
\end{eqnarray}%
where the coefficients are%
\begin{eqnarray}
a_{d} &=&\sum\limits_{i=0}^{\min \left( N_{1}-1,N_{2}\right) }a_{d}\left(
i\right) ,  \label{ad} \\
b_{d} &=&\sum\limits_{i=0}^{\min \left( N_{1}-1,N_{2}-1\right) }b_{d}\left(
i\right) ,  \label{bd} \\
c_{d} &=&\sum\limits_{i=0}^{\min \left( N_{1},N_{2}-1\right) }c_{d}\left(
i\right) .  \label{cd}
\end{eqnarray}%
In the above summations,%
\begin{eqnarray}
a_{d}\left( i\right) &=&\frac{A_{n}^{2}N_{1}!N_{2}!\left\vert \zeta \left(
t\right) \right\vert ^{2i}}{i!i!\left( N_{1}-i-1\right) !\left(
N_{2}-i\right) !},  \label{adi} \\
b_{d}\left( i\right) &=&\frac{A_{n}^{2}N_{1}!N_{2}!\left\vert \zeta \left(
t\right) \right\vert ^{2i+1}}{i!\left( i+1\right) !\left( N_{1}-i-1\right)
!\left( N_{2}-i-1\right) !},  \label{bdi} \\
c_{d}\left( i\right) &=&\frac{A_{n}^{2}N_{1}!N_{2}!\left\vert \zeta \left(
t\right) \right\vert ^{2i}}{i!i!\left( N_{1}-i\right) !\left(
N_{2}-i-1\right) !}.  \label{cdi}
\end{eqnarray}

For two independent ideal condensates, before the overlapping between the
two condensates, we have $\zeta \left( t=0\right) =0$. Based on the Schr\H{o}%
dinger equation, it is easy to verify that after the double-well potential
separating the condensates is removed, we have $\zeta \left( t\right) =0$ at
any further time. Thus $b_{d}=0$, and the density expectation value is given
by%
\begin{equation}
n_{d}\left( \mathbf{r},t\right) =N_{1}\left\vert \phi _{1}\left( \mathbf{r}%
,t\right) \right\vert ^{2}+N_{2}\left\vert \phi _{2}\left( \mathbf{r}%
,t\right) \right\vert ^{2}.
\end{equation}%
In this situation, the interference term is zero in the density expectation
value.

In the presence of the interatomic interaction, $\zeta \left( t\right) $ can
be a nonzero value after the two initially independent condensates begin to
overlap. For $\zeta \left( t\right) $ being nonzero, we see clearly from the
coefficient $b_{d}$ that the interference term in (\ref{density}) gives a
contribution to the density expectation value. Based on Eqs. (\ref{adi}), (%
\ref{bdi}) and (\ref{cdi}), we have%
\begin{equation}
\frac{b_{d}\left( i\right) }{c_{d}\left( i\right) }=\frac{\left\vert \zeta
\left( t\right) \right\vert \left( N_{1}-i\right) }{i+1},
\end{equation}%
and%
\begin{equation}
\frac{b_{d}\left( i\right) }{a_{d}\left( i\right) }=\frac{\left\vert \zeta
\left( t\right) \right\vert \left( N_{2}-i\right) }{i+1}.
\end{equation}%
It is easy to understand that when $N_{1}\left\vert \zeta \left( t\right)
\right\vert >1$ and $N_{2}\left\vert \zeta \left( t\right) \right\vert >1$, $%
b_{d}$ can not be omitted, and thus there would be clear interference
patterns. This shows clearly that for large particle number, a small $%
\left\vert \zeta \left( t\right) \right\vert $ can play important role in
the density expectation value when $\left\vert \zeta \left( t\right)
\right\vert >N_{1}^{-1}$ and $\left\vert \zeta \left( t\right) \right\vert
>N_{2}^{-1}$.

Generally speaking, even in the presence of the interatomic interaction, $%
\left\vert \zeta \left( t\right) \right\vert $ is much smaller than $1$
because $\phi _{1}\phi _{2}^{\ast }$ is an oscillation function about the
space coordinate. As shown above, however, a nonzero value of $\zeta \left(
t\right) $ can give significant contribution to the density expectation
value for large $N_{1}$ and $N_{2}$. In Fig. 2(a), we give the relation
between $b_{d}/c_{d}$ and $\zeta $ for $N_{1}=N_{2}=10^{3}$ based on Eqs. (%
\ref{bd}) and (\ref{cd}). The relation between $b_{d}/c_{d}$ and $%
N_{1}=N_{2} $ for $\zeta =0.001$ is shown in Fig. 2(b). Generally speaking,
for $N_{1}\left\vert \zeta \right\vert >>1$ and $N_{2}\left\vert \zeta
\right\vert >>1$, one has $b_{d}/\sqrt{a_{d}c_{d}}\approx 1$.

\begin{figure}[tbp]
\includegraphics[width=0.8\linewidth,angle=270]{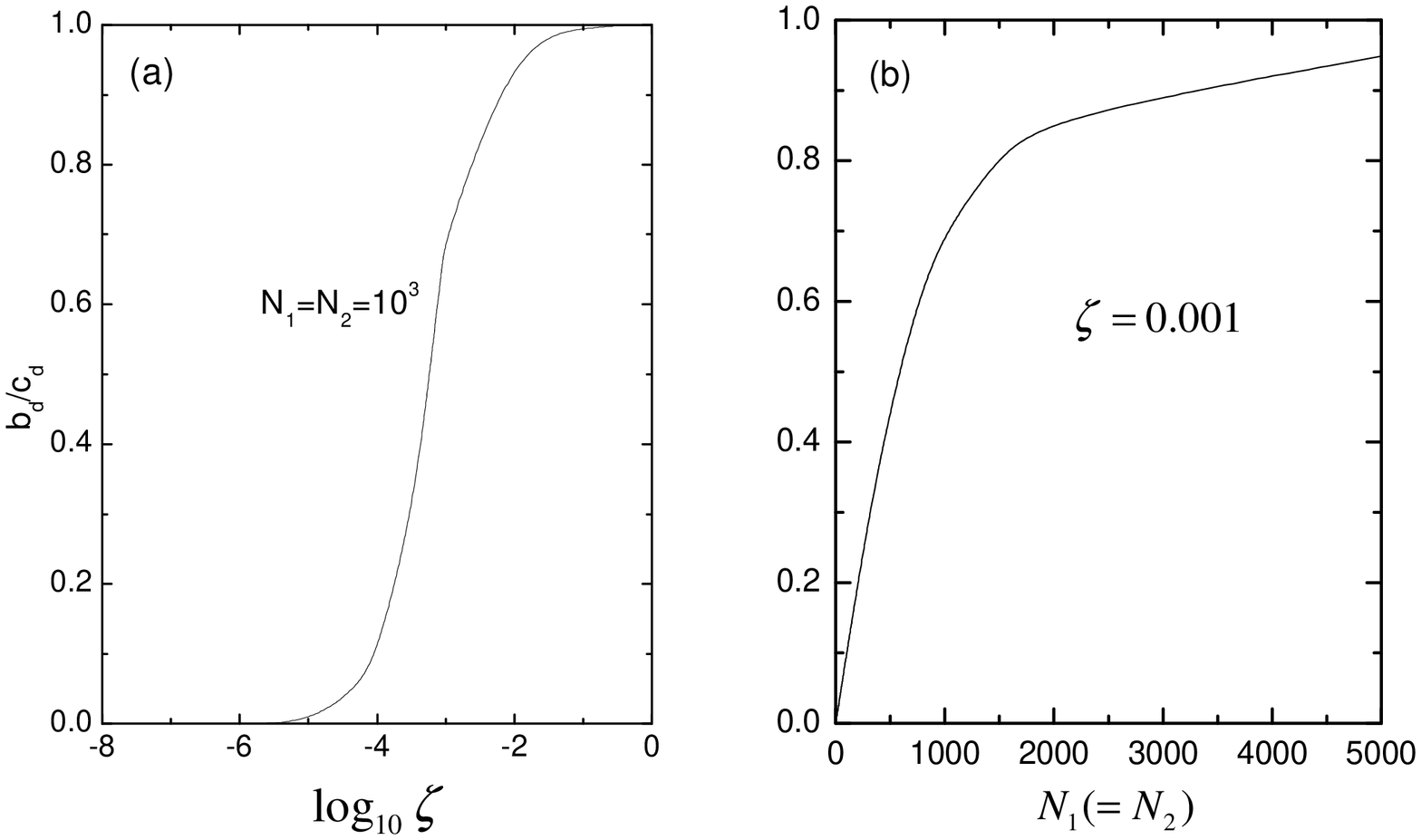}
\caption{Based on the numerical calculations of Eqs. (\protect\ref{ad}), (%
\protect\ref{bd}) and (\protect\ref{cd}), shown in Fig. 2(a) is the relation
between $b_{d}/c_{d}$ and $\protect\zeta $, while shown in Fig. 2(b) is the
relation between $b_{d}/c_{d}$ and $N_{1}=N_{2}$ for $\protect\zeta =0.001$.
It is shown clearly that the interference term can play an important role in
the density expectation value for $\left\vert \protect\zeta \right\vert $
being larger than $N_{1}^{-1}$.}
\end{figure}

\section{the evolution equations of the system}

For the many-body wave function $\Psi _{N_{1}N_{2}}$, in the presence of the
interatomic interaction, the evolution equation can be obtained based on the
standard quantum mechanical principle. After the double-well potential is
removed, the evolution equation is given by%
\begin{equation}
i\hbar \frac{\partial \Psi _{N_{1}N_{2}}}{\partial t}=\widehat{H}_{f}\Psi
_{N_{1}N_{2}},  \label{evolution-hamiltonian}
\end{equation}%
where the Hamiltonian of the whole system in the first-quantization method is%
\begin{equation}
\widehat{H}_{f}=\sum_{i=1}^{N}-\frac{\hslash ^{2}}{2m}\nabla
_{i}^{2}+g\sum_{i<j}^{N}\delta \left( \mathbf{r}_{i}-\mathbf{r}_{j}\right) .
\label{hamiltonian-first}
\end{equation}%
Here the coupling constant $g=4\pi \hbar ^{2}a_{s}/m$ with $a_{s}$ being the
scattering length. In the above expression of the Hamiltonian, we have used
the well-known two-body pseudopotentials.

It is well known that the action principle is quite useful to derive the GP
equation for a single condensate (See for example Ref. \cite{Pethick}).
Similarly, we consider here the evolution of $\Psi _{N_{1}N_{2}}$ based on
the action principle. To get the evolution equation based on the action
principle, we first give the general expression of the overall energy of the
whole system. After removing the double-well potential, the overall energy
of the whole system is
\begin{equation}
E=\int \Psi _{N_{1}N_{2}}^{\ast }\widehat{H}_{f}\Psi _{N_{1}N_{2}}d^{3}%
\mathbf{r}_{1}\cdots d^{3}\mathbf{r}_{N}.  \label{overallenergy}
\end{equation}%
After straightforward derivations, the exact expression of the overall
energy is given by%
\begin{equation}
E=E_{kin}+E_{int},  \label{S-overallenergy}
\end{equation}%
where the kinetic energy $E_{kin}$ is%
\begin{eqnarray}
E_{kin} &=&\int \Psi _{N_{1}N_{2}}^{\ast }\left( \sum_{i=1}^{N}-\frac{%
\hslash ^{2}}{2m}\nabla _{i}^{2}\right) \Psi _{N_{1}N_{2}}d^{3}\mathbf{r}%
_{1}\cdots d^{3}\mathbf{r}_{N}  \nonumber \\
&=&\int dV\left( \frac{a_{d}\hslash ^{2}}{2m}\nabla \phi _{1}^{\ast }\cdot
\nabla \phi _{1}+\frac{b_{d}\hbar ^{2}}{2m}e^{i\varphi _{c}}\nabla \phi
_{1}^{\ast }\cdot \nabla \phi _{2}\right.  \nonumber \\
&&\left. +\frac{b_{d}\hbar ^{2}}{2m}e^{-i\varphi _{c}}\nabla \phi _{2}^{\ast
}\cdot \nabla \phi _{1}+\frac{c_{d}\hbar ^{2}}{2m}\nabla \phi _{2}^{\ast
}\cdot \nabla \phi _{2}\right) .
\end{eqnarray}%
In addition, the interaction energy $E_{int}$ of the whole system is given by%
\begin{eqnarray}
E_{int} &=&\int \Psi _{N_{1}N_{2}}^{\ast }\left( g\sum_{i<j}^{N}\delta
\left( \mathbf{r}_{i}-\mathbf{r}_{j}\right) \right) \Psi _{N_{1}N_{2}}d^{3}%
\mathbf{r}_{1}\cdots d^{3}\mathbf{r}_{N}  \nonumber \\
&=&\frac{g}{2}\int dV\left[ h_{1}\left\vert \phi _{1}\right\vert
^{4}+h_{2}\left\vert \phi _{2}\right\vert ^{4}+h_{3}\left\vert \phi
_{1}\right\vert ^{2}\left\vert \phi _{2}\right\vert ^{2}\right.  \nonumber \\
&&\left. +\mathrm{Re}\left( h_{4}\left\vert \phi _{1}\right\vert ^{2}\phi
_{1}^{\ast }\phi _{2}e^{i\varphi _{c}}+h_{5}\left( \phi _{1}^{\ast }\right)
^{2}\phi _{2}^{2}e^{2i\varphi _{c}}+h_{6}\left\vert \phi _{2}\right\vert
^{2}\phi _{1}^{\ast }\phi _{2}e^{i\varphi _{c}}\right) \right] ,
\label{exact-inter}
\end{eqnarray}%
where the coefficients are given by%
\begin{eqnarray}
h_{1} &=&\sum_{i=0}^{\min \left( N_{1}-2,N_{2}\right) }\frac{%
A_{n}^{2}N_{1}!N_{2}!\left\vert \zeta \right\vert ^{2i}}{i!i!\left(
N_{1}-i-2\right) !\left( N_{2}-i\right) !},  \nonumber \\
h_{2} &=&\sum_{i=0}^{\min \left( N_{1},N_{2}-2\right) }\frac{%
A_{n}^{2}N_{1}!N_{2}!\left\vert \zeta \right\vert ^{2i}}{i!i!\left(
N_{1}-i\right) !\left( N_{2}-i-2\right) !},  \nonumber \\
h_{3} &=&\sum_{i=0}^{\min \left( N_{1}-1,N_{2}-1\right) }\frac{%
4A_{n}^{2}N_{1}!N_{2}!\left\vert \zeta \right\vert ^{2i}}{i!i!\left(
N_{1}-i-1\right) !\left( N_{2}-i-1\right) !},  \nonumber \\
h_{4} &=&\sum_{i=0}^{\min \left( N_{1}-2,N_{2}-1\right) }\frac{%
4A_{n}^{2}N_{1}!N_{2}!\left\vert \zeta \right\vert ^{2i+1}}{i!\left(
i+1\right) !\left( N_{1}-i-2\right) !\left( N_{2}-i-1\right) !},  \nonumber
\\
h_{5} &=&\sum_{i=0}^{\min \left( N_{1}-2,N_{2}-2\right) }\frac{%
2A_{n}^{2}N_{1}!N_{2}!\left\vert \zeta \right\vert ^{2i+2}}{i!\left(
i+2\right) !\left( N_{1}-i-2\right) !\left( N_{2}-i-2\right) !},  \nonumber
\\
h_{6} &=&\sum_{i=0}^{\min \left( N_{1}-1,N_{2}-2\right) }\frac{%
4A_{n}^{2}N_{1}!N_{2}!\left\vert \zeta \right\vert ^{2i+1}}{i!\left(
i+1\right) !\left( N_{1}-i-1\right) !\left( N_{2}-i-2\right) !}.
\label{alpha-coe}
\end{eqnarray}

By using the ordinary action principle and the energy of the whole system,
one can get the following coupled evolution equations for $\phi _{1}$ and $%
\phi _{2}$:
\begin{eqnarray}
i\hslash \frac{\partial \phi _{1}}{\partial t} &=&\frac{1}{N_{1}}\frac{%
\delta E}{\delta \phi _{1}^{\ast }},  \label{GP1} \\
i\hslash \frac{\partial \phi _{2}}{\partial t} &=&\frac{1}{N_{2}}\frac{%
\delta E}{\delta \phi _{2}^{\ast }},  \label{GP2}
\end{eqnarray}%
where $\delta E/\delta \phi _{1}^{\ast }$ and $\delta E/\delta \phi
_{2}^{\ast }$ are functional derivatives.

\section{the crossover from $\left\vert \protect\zeta \right\vert <<N^{-1}$
to $\left\vert \protect\zeta \right\vert >>N^{-1}$}

\subsection{the case of $\left\vert \protect\zeta \right\vert <<N^{-1}$}

In Sec. II, we have shown that the nonzero interference term in the density
expectation value origins from the assumption that $\zeta \left( t\right) $
can be a nonzero value after the overlapping between two initially
independent condensates. In the last section, we have given the evolution
equations about $\phi _{1}$ and $\phi _{2}$. Thus, an important question
emerges naturally: is it physical for $\zeta \left( t\right) $ being nonzero
with the development of time?

Based on the evolution equations (\ref{GP1}) and (\ref{GP2}) given in the
last section, one can understand easily that $\zeta \left( t\right) $
becomes nonzero after the overlapping between two initially independent
condensates in the presence of the interatomic interaction. Before removing
the double-well potential, $N_{1}\left\vert \zeta \right\vert =0$ and $%
N_{2}\left\vert \zeta \right\vert =0$ because there is no overlapping
between $\phi _{1}$ and $\phi _{2}$. After removing the double-well
potential and at the beginning of the overlapping between two condensates, $%
N_{1}\left\vert \zeta \right\vert <<1$ and $N_{2}\left\vert \zeta
\right\vert <<1$. For $N_{1}\left\vert \zeta \right\vert <<1$ and $%
N_{2}\left\vert \zeta \right\vert <<1$, the overall energy can be
approximated as%
\begin{eqnarray}
E &\approx &\frac{\hslash ^{2}}{2m}\int dV\left( N_{1}\nabla \phi _{1}^{\ast
}\cdot \nabla \phi _{1}+N_{2}\nabla \phi _{2}^{\ast }\cdot \nabla \phi
_{2}\right) +  \nonumber \\
&&\frac{g}{2}\int dV\left( N_{1}\left( N_{1}-1\right) \left\vert \phi
_{1}\right\vert ^{4}+N_{2}\left( N_{2}-1\right) \left\vert \phi
_{2}\right\vert ^{4}+4N_{1}N_{2}\left\vert \phi _{1}\right\vert
^{2}\left\vert \phi _{2}\right\vert ^{2}\right) .  \label{app-energy-small}
\end{eqnarray}%
The above overall energy is obtained by setting $\zeta =0$ in Eq. (\ref%
{S-overallenergy}).

Based on Eqs. (\ref{GP1}) and (\ref{GP2}), the approximate evolution
equations for $\phi _{1}\left( \mathbf{r},t\right) $ and $\phi _{2}\left(
\mathbf{r},t\right) $ are then
\begin{equation}
i\hslash \frac{\partial \phi _{1}}{\partial t}=-\frac{\hslash ^{2}}{2m}%
\nabla ^{2}\phi _{1}+V_{1}\phi _{1},  \label{S-GP1}
\end{equation}%
and
\begin{equation}
i\hslash \frac{\partial \phi _{2}}{\partial t}=-\frac{\hslash ^{2}}{2m}%
\nabla ^{2}\phi _{2}+V_{2}\phi _{2},  \label{S-GP2}
\end{equation}%
where
\begin{equation}
V_{1}=\left( N_{1}-1\right) g\left\vert \phi _{1}\right\vert
^{2}+2N_{2}g\left\vert \phi _{2}\right\vert ^{2},
\end{equation}%
and
\begin{equation}
V_{2}=\left( N_{2}-1\right) g\left\vert \phi _{2}\right\vert
^{2}+2N_{1}g\left\vert \phi _{1}\right\vert ^{2}.
\end{equation}

We see that $V_{1}$ is not equal to $V_{2}$ for $g\neq 0$. This leads to an
important result that $\zeta \left( t\right) $ can be nonzero when there is
an overlapping between two initially independent condensates. One can
understand this result further through the following equation which
determines the evolution of $\zeta \left( t\right) $:
\begin{equation}
i\hslash \frac{d\zeta \left( t\right) }{dt}=\int f\left( \mathbf{r},t\right)
\phi _{1}\phi _{2}^{\ast }dV,  \label{S-xi-t}
\end{equation}%
where $f\left( \mathbf{r},t\right) $ is a nonzero function given by
\begin{equation}
f\left( \mathbf{r},t\right) =\left( N_{2}+1\right) g\left\vert \phi
_{2}\right\vert ^{2}-\left( N_{1}+1\right) g\left\vert \phi _{1}\right\vert
^{2}.  \label{S-f-function}
\end{equation}

The above analyses show clearly that why $\zeta \left( t\right) $ becomes
nonzero after the overlapping between two initially independent condensates
for $g\neq 0$. Obviously, for $g$ being zero, $\zeta \left( t\right) =0$ at
any further time because $d\zeta \left( t\right) /dt=0$. Our numerical
calculations in the following section also show that $\zeta \left( t\right) $
can be a nonzero value in the presence of the interatomic interaction.

In the presence of the interatomic interaction, it is inconsistent to assume
that $\zeta \left( t\right) $ is always zero with the development of time.
If $\zeta \left( t\right) =0$, the evolution equations Eqs. (\ref{GP1}) and (%
\ref{GP2}) are exact. However, based on these two evolution equations, $%
\left\vert \zeta \right\vert $ will increase from zero after the overlapping.

\subsection{the case of $\left\vert \protect\zeta \right\vert >>N^{-1}$}

As shown above, the evolution of $\phi _{1}$ and $\phi _{2}$ is determined
by the expression of the overall energy and the evolution equations (\ref%
{GP1}) and (\ref{GP2}). It seems that there is no exact solution for these
evolution equations considering the fact that even there is no exact
solution for three-dimensional nonlinear Schr\H{o}dinger equation. However,
for $\left\vert \zeta \right\vert >>N_{1}^{-1}$ and $\left\vert \zeta
\right\vert >>N_{2}^{-1}$, we find that there is a quite simple evolution
equation by introducing an effective parameter order. This would also
contribute to our understanding of the interaction-induced coherence process
for two initially independent condensates.

For the cases of $N_{1}\left\vert \zeta \right\vert >>1$, $N_{2}\left\vert
\zeta \right\vert >>1$ and $N_{1}\sim N_{2}$, first we introduce the
following effective order parameter $\Phi _{e}\left( \mathbf{r},t\right) $
which is given by
\begin{equation}
\Phi _{e}\left( \mathbf{r},t\right) =\sqrt{N_{1}}\phi _{1}\left( \mathbf{r}%
,t\right) +\sqrt{N_{2}}e^{i\varphi _{c}}\phi _{2}\left( \mathbf{r},t\right) .
\end{equation}%
Based on this effective order parameter, the density expectation value can
be approximated well as%
\begin{eqnarray}
n_{d}\left( \mathbf{r},t\right) &\simeq &\Phi _{e}^{\ast }\left( \mathbf{r}%
,t\right) \Phi _{e}\left( \mathbf{r},t\right)  \nonumber \\
&=&a_{d}^{\prime }\left\vert \phi _{1}\left( \mathbf{r},t\right) \right\vert
^{2}+2b_{d}^{\prime }\times \mathrm{Re}\left[ e^{i\varphi _{c}}\phi
_{1}^{\ast }\left( \mathbf{r},t\right) \phi _{2}\left( \mathbf{r},t\right) %
\right] +c_{d}^{\prime }\left\vert \phi _{2}\left( \mathbf{r},t\right)
\right\vert ^{2},  \label{app-density}
\end{eqnarray}%
where%
\begin{eqnarray}
a_{d}^{\prime } &=&N_{1},  \nonumber \\
b_{d}^{\prime } &=&\sqrt{N_{1}N_{2}},  \nonumber \\
c_{d}^{\prime } &=&N_{2}.
\end{eqnarray}

To compare with the exact expression of the density expectation value given
by Eq. (\ref{density}), Fig. 3 shows the ratio $\lambda
_{1}=a_{d}/a_{d}^{\prime }$, $\lambda _{2}=b_{d}/b_{d}^{\prime }$ and $%
\lambda _{3}=c_{d}/c_{d}^{\prime }$. We see that for $N_{1}\left\vert \zeta
\right\vert >>1$, $N_{2}\left\vert \zeta \right\vert >>1$, the approximate
density expectation value $\Phi _{e}^{\ast }\left( \mathbf{r},t\right) \Phi
_{e}\left( \mathbf{r},t\right) $ agrees very well with the exact expression
of the density expectation value.

\begin{figure}[tbp]
\includegraphics[width=0.8\linewidth,angle=270]{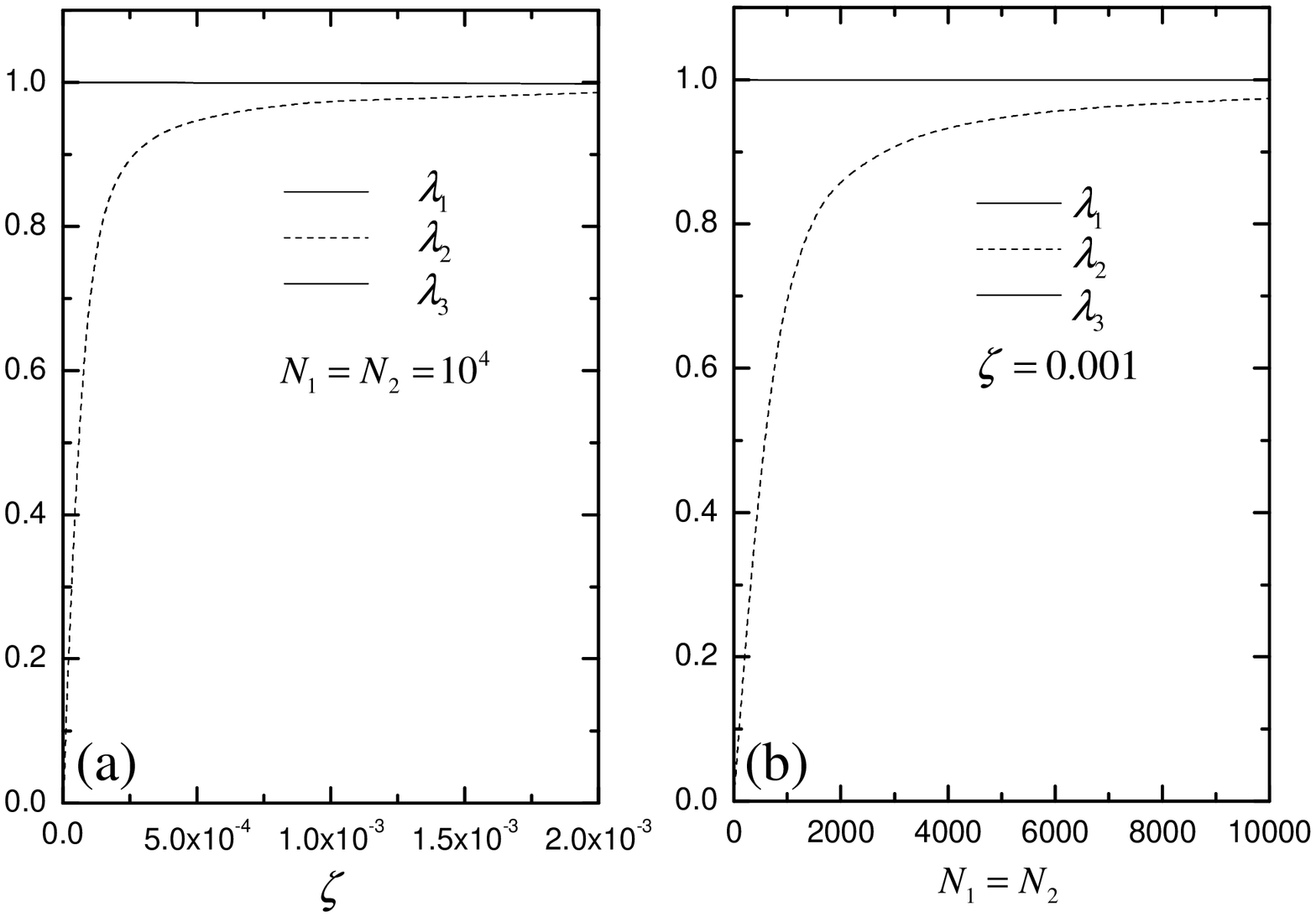}
\caption{Shown is the values of $\protect\lambda _{1}$, $\protect\lambda %
_{2} $ and $\protect\lambda _{3}$. We see that for $N_{1}\left\vert \protect%
\zeta \right\vert >>1$, $N_{2}\left\vert \protect\zeta \right\vert >>1$ and $%
N_{1}\sim N_{2}$, the density expectation value given by the effective order
parameter agrees well with the exact expression of the density expectation
value given by Eq. (\protect\ref{density}).}
\end{figure}

Based on the effective order parameter, for $N_{1}\left\vert \zeta
\right\vert >>1$, $N_{2}\left\vert \zeta \right\vert >>1$ and $N_{1}\sim
N_{2}$, we also find that the overall energy of the whole system can be
approximated very well as
\begin{equation}
E_{app}^{\prime }=E_{kin}^{\prime }+E_{int}^{\prime },  \label{S-app-energy}
\end{equation}%
where%
\begin{equation}
E_{kin}^{\prime }=\frac{\hbar ^{2}}{2m}\int \nabla \Phi _{e}^{\ast }\cdot
\nabla \Phi _{e}dV,  \label{appro-inter}
\end{equation}%
and%
\begin{equation}
E_{int}^{\prime }=\frac{g}{2}\int dV\left\vert \Phi _{e}\right\vert ^{4}.
\label{eappr-inter}
\end{equation}

For the cases of $N_{1}\left\vert \zeta \right\vert >>1$, $N_{2}\left\vert
\zeta \right\vert >>1$ and $N_{1}\sim N_{2}$, it is easy to verify that $%
E_{kin}^{\prime }\approx E_{kin}$ based on the analogous analyses about the
density expectation value. For $N_{1}\left\vert \zeta \right\vert >>1$, $%
N_{2}\left\vert \zeta \right\vert >>1$ and $N_{1}\sim N_{2}$, one can also
prove the result of $E_{int}^{\prime }\approx E_{int}$. Based on Eq. (\ref%
{eappr-inter}), $E_{int}^{\prime }$ can be expanded as:
\begin{eqnarray}
E_{int}^{\prime } &=&\frac{g}{2}\int dV\left[ \beta _{1}\left\vert \phi
_{1}\right\vert ^{4}+\beta _{2}\left\vert \phi _{2}\right\vert ^{4}+\beta
_{3}\left\vert \phi _{1}\right\vert ^{2}\left\vert \phi _{2}\right\vert
^{2}\right.  \nonumber \\
&&\left. +\mathrm{Re}\left( \beta _{4}\left\vert \phi _{1}\right\vert
^{2}\phi _{1}^{\ast }\phi _{2}e^{i\varphi _{c}}+\beta _{5}\left( \phi
_{1}^{\ast }\right) ^{2}\phi _{2}^{2}e^{2i\varphi _{c}}+\beta _{6}\left\vert
\phi _{2}\right\vert ^{2}\phi _{1}^{\ast }\phi _{2}e^{i\varphi _{c}}\right) %
\right] ,
\end{eqnarray}%
where
\begin{eqnarray}
\beta _{1} &=&N_{1}^{2},  \nonumber \\
\beta _{2} &=&N_{2}^{2},  \nonumber \\
\beta _{3} &=&4N_{1}N_{2},  \nonumber \\
\beta _{4} &=&4N_{1}\sqrt{N_{1}N_{2}},  \nonumber \\
\beta _{5} &=&2N_{1}N_{2},  \nonumber \\
\beta _{6} &=&4N_{2}\sqrt{N_{1}N_{2}}.
\end{eqnarray}%
To compare with the exact expression of the interaction energy given by Eq. (%
\ref{exact-inter}), Fig. 4 shows the relation between $h_{i}/\beta _{i}$ ($%
i=1,\cdots ,6$) and $\zeta $ for $N_{1}=N_{2}=10^{3}$. It is shown clearly
that for $N_{1}\left\vert \zeta \right\vert >>1$ and $N_{2}\left\vert \zeta
\right\vert >>1$, $h_{i}/\beta _{i}\approx 1$, and thus $E_{int}^{\prime
}\approx E_{int}$.

In this situation, the overall energy can be approximated well as
\begin{equation}
E_{app}^{\prime }=\frac{\hbar ^{2}}{2m}\int \nabla \Phi _{e}^{\ast }\cdot
\nabla \Phi _{e}dV+\frac{g}{2}\int dV\left\vert \Phi _{e}\right\vert ^{4}.
\end{equation}%
The evolution of the effective order parameter can be then obtained by using
the action principle. Based on this approximate energy, after removing the
double-well potential, it is quite interesting to note that the evolution of
the effective order parameter can be described very well by the ordinary
Gross-Pitaevskii equation:
\begin{equation}
i\hslash \frac{\partial \Phi _{e}}{\partial t}\simeq -\frac{\hslash ^{2}}{2m}%
\nabla ^{2}\Phi _{e}+g\left\vert \Phi _{e}\right\vert ^{2}\Phi _{e}.
\label{S-app-equation}
\end{equation}%
\begin{figure}[tbp]
\includegraphics[width=0.8\linewidth,angle=270]{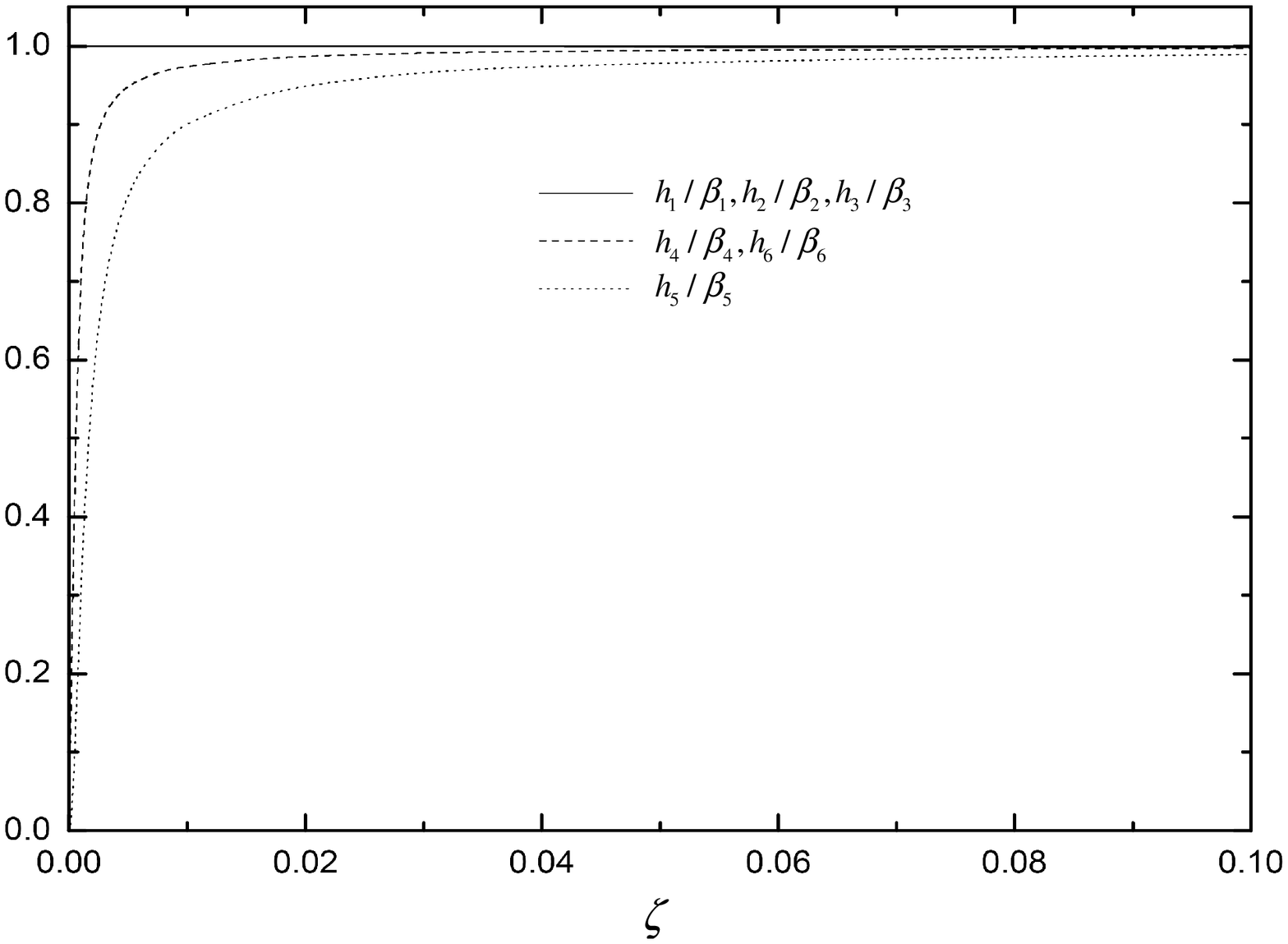}
\caption{Shown is the relation between $h_{i}/\protect\beta _{i}$ ($%
i=1,\cdots ,6$) and $\protect\zeta $ for $N_{1}=N_{2}=10^{3}$. We see that
for $N_{1}\left\vert \protect\zeta \right\vert >>1$ and $N_{2}\left\vert
\protect\zeta \right\vert >>1$, $h_{i}/\protect\beta _{i}\approx 1$, which
means that $E_{int}^{\prime }\approx E_{int}$.}
\end{figure}

Based on Eqs. (\ref{app-density}) and (\ref{S-app-equation}), we see that
the emergence of the effective order parameter $\Phi _{e}$ gives us strong
evidence that the coherence is formed in the interaction process between two
initially independent condensates, and thus results in the emergence of
high-contrast interference fringes. The effective order parameter and the
approximate density expectation value suggest strongly that a full coherence
is formed between two initially independent condensates for the cases of $%
N_{1}\left\vert \zeta \right\vert >>1$ and $N_{2}\left\vert \zeta
\right\vert >>1$.

\section{the evolution of the density expectation value according to the
experimental parameters}

Now we turn to give the theoretical results of the density expectation value
according to the experimental parameters in Ref. \cite{Andrew} where clear
interference patterns were observed for two initially independent
condensates. In the experiment of Ref. \cite{Andrew}, $N=5\times 10^{6}$
condensed sodium atoms were confined in a magnetic trap with $\omega
_{x}=2\pi \times 18$ \textrm{Hz}, $\omega _{y}=\omega _{z}=2\pi \times 320$
\textrm{Hz}. A blue-detuned laser beam of wavelength $514$ \textrm{nm} was
focused into a light sheet with a cross section of $12$ $\mathrm{\mu m}$ by $%
67$ $\mathrm{\mu m}$. The long axis of the laser beam was perpendicular to
the long $x-$axis of the condensate. For a laser power of $14$ \textrm{mW},
the barrier height is about $1.4$ $\mathrm{\mu K}$, which is much larger
than the chemical potential $\mu =0.03$ $\mathrm{\mu K}$. For this laser
beam, the two condensates can be regarded to be independent because they are
well separated and the tunneling effect can be omitted. With these
experimental parameters and the $s-$wave scattering length $a_{s}=2.75$
\textrm{nm}, the initial profile of the two condensates is shown in Fig. 5.
At the initial time, the overlapping between two condensates can be omitted
safely, and thus $\zeta \left( t=0\right) =0$. After the double-well
potential is removed, the evolution of the density expectation value $%
n_{d-x}\left( x,t\right) =\int n_{d}\left( \mathbf{r},t\right) dydz$ (in
unit of $N/2$) is given in Fig. 5 through the numerical calculations of Eqs.
(\ref{GP1}), (\ref{GP2}) and (\ref{density}). We see that there is a clear
interference pattern in the density expectation value which agrees with the
experimental result. Shown in the inset is the evolution of $\left\vert
\zeta \right\vert $ for these parameters. We see that $\left\vert \zeta
\right\vert $ increases from zero with the development of time. For the
expansion time of $40$ \textrm{ms}, the numerical result of $\left\vert
\zeta \right\vert $ shows that $b_{d}/c_{d}\approx 1$. To check further that
the nonzero value of $\left\vert \zeta \right\vert $ does not originate from
numerical error, with the same initial conditions, we have verified in the
numerical calculations that $\left\vert \zeta \right\vert $ is always zero
(smaller than $10^{-10}$) if the scattering length is assumed as zero.

\begin{figure}[tbp]
\includegraphics[width=0.8\linewidth,angle=270]{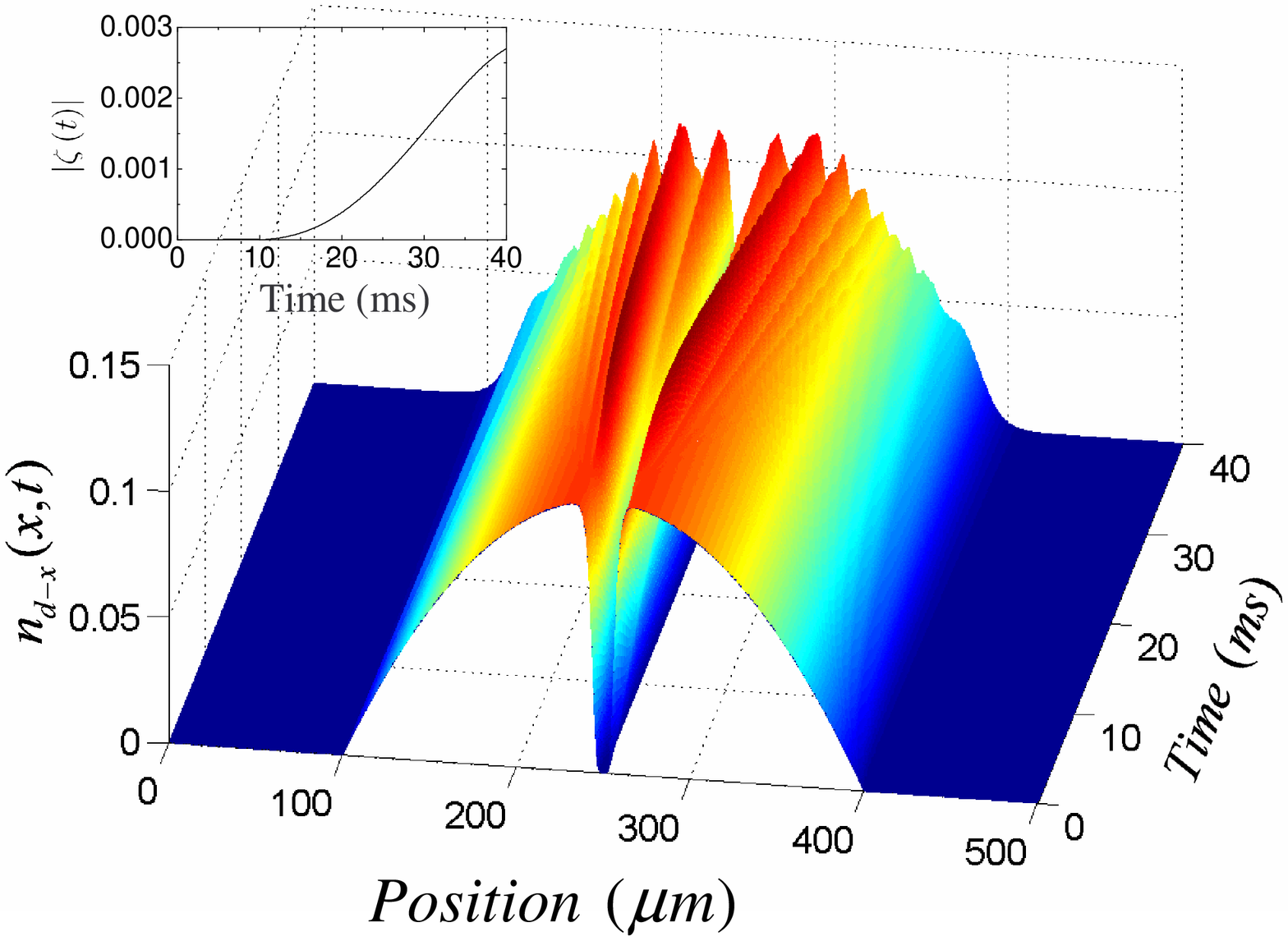}
\caption{Shown is the evolution of the density expectation value calculated
using the experimental parameters in Ref. \protect\cite{Andrew}. It is shown
that there is a clear interference pattern after the overlapping between two
independent interacting condensates. Shown in the inset is the evolution of $%
\left\vert \protect\zeta \left( t\right) \right\vert $ for the experimental
parameters in Ref. \protect\cite{Andrew}.}
\end{figure}

Based on the evolution of the density expectation value shown in Fig. 5, we
see that the overall width (about $300$ $\mathrm{\mu m}$) of the ultracold
gases in the $x-$direction does not increase obviously. This is due to the
fact that the initial density distribution is cigar-shaped, and thus the
expansion in the $x-$direction is very slow, while the expansion in $y$ and $%
z$ directions is much quick. In the experiment of Ref. \cite{cigar}, one can
see clearly that there is no obvious expansion in the long $x-$axis for
cigar-shaped condensate. For two initially independent condensates, when the
double-well potential is switched off, one should note that in the regime
close to $x=250$ $\mathrm{\mu m}$ shown in Fig. 5, the ultracold gases
expand rapidly in the $x-$direction because in this regime the ultracold
gases have higher kinetic energy. Thus, although the total width of the
system in the $x-$direction does not increase obviously, the rapid expansion
in the central regime leads to the overlapping between two initially
independent condensates, and results in the interference effect. The overall
width of the system shown in Fig. 5 is smaller than the experimental result
of about $500$ $\mathrm{\mu m}$ \cite{Andrew}. This difference may come from
the expansion of thermal cloud in this experiment \cite{Andrew}. After $40$
\textrm{ms} expansion, the numerical result in \cite{Rohrl} for two
coherently separated condensates also showed that the overall width of the
system is about $300$ $\mathrm{\mu m}$.

\section{the evolution of the density expectation value for different
coupling constants}

\bigskip We see that the interatomic interaction plays an essential role in
the emergence of the interference effect for two initially independent
condensates. Generally speaking, increasing the particle number will enhance
the effect of the interference term in the density expectation value. Based
on Eqs. (\ref{GP1}) and (\ref{GP2}), increasing the coupling constant $g$\
has the effect of increasing $\zeta \left( t\right) $. Together with the
relation between $b_{d}/c_{d}$\ and $\zeta $\ illustrated in Fig. 2(a), this
shows that increasing the interatomic interaction will enhance the effect of
the interference term. To show more clearly the interaction-induced
coherence process between two initially independent condensates, in this
section we consider the density expectation value for different coupling
constants.

We consider here the evolution of the density expectation value for
one-dimensional case. At $t=0$, to give a general comparison, the initial
wave functions for two independent condensates are assumed to be identical
for different coupling constants. The initial wave functions are
respectively given by
\begin{eqnarray}
\phi _{1}\left( x_{l},t=0\right) &=&\frac{1}{\pi ^{1/4}\sqrt{\Delta _{1}}}%
\exp \left[ -\frac{\left( x_{l}-x_{l1}\right) ^{2}}{2\Delta _{1}^{2}}\right]
,  \label{inital1} \\
\phi _{2}\left( x_{l},t=0\right) &=&\frac{1}{\pi ^{1/4}\sqrt{\Delta _{2}}}%
\exp \left[ -\frac{\left( x_{l}-x_{l2}\right) ^{2}}{2\Delta _{2}^{2}}\right]
.  \label{inital2}
\end{eqnarray}%
In the above wave functions, we have introduced a dimensionless variable $%
x_{l}=x/l$ with $l$ being a length. In the present work, we assume that $%
\Delta _{1}=\Delta _{2}=0.5$ and $x_{l2}-x_{l1}=4.5$. For these parameters,
at $t=0$, the two condensates are well separated. In the numerical
calculations of the coupled equations given by Eqs. (\ref{GP1}) and (\ref%
{GP2}), it is useful to introduce the dimensionless variable $\tau
=E_{l}t/\hslash $ with $E_{l}=\hbar ^{2}/2ml^{2}$, and dimensionless
coupling constant $g_{l}=N_{1}g/E_{l}l$.\ In addition, the particle number
is assumed as $N_{1}=N_{2}=1.0\times 10^{5}$. In real experiments,
interatomic interaction plays very important role in the initial
ground-state wave function of the condensates. However, in principle, one
can prepare the state given by Eqs. (\ref{inital1}) and (\ref{inital2}) by
adjusting the trapping potential for different coupling constants. In this
section, the identical initial wave functions for different coupling
constants would be helpful in the comparison of the density expectation
value for different coupling constants.

For $t>0$, we consider the evolution of the density expectation value in
free space. The evolution of $\phi _{1}$ and $\phi _{2}$ is obtained based
on the numerical calculations of Eqs. (\ref{GP1}) and (\ref{GP2}). From $%
\phi _{1}$ and $\phi _{2}$, we can get $\zeta $, and thus the density
expectation value based on Eq. (\ref{density}). Shown in Fig. 6 is the
evolution of the density expectation value for different coupling constants.
It is shown clearly that increasing the coupling constant has the effect of
enhancing the coherence effect, and results in higher contrast in the
interference patterns.

\begin{figure}[tbp]
\includegraphics[width=0.8\linewidth,angle=270]{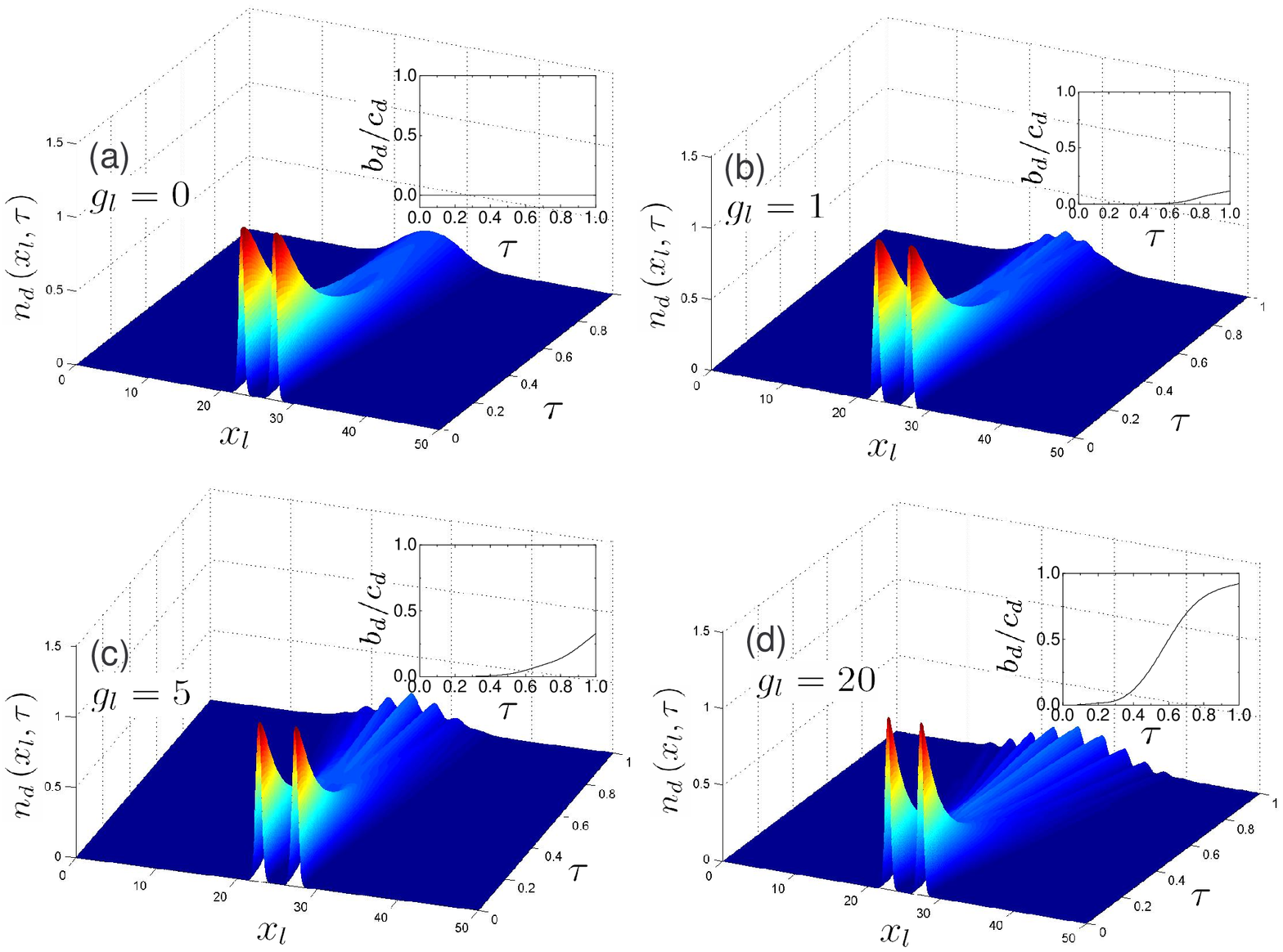}
\caption{After two initially independent condensates are allowed to expand
freely, shown is the evolution of the density expectation value $n_{d}\left(
x_{l},\protect\tau \right) $ (in unit of $N_{1}+N_{2}$) for different
coupling constants $g_{l}$. Shown in the inset of each figure is the
relation between $b_{d}/c _{d}$ and dimensionless time $\protect\tau $. For
two ideal condensates shown in Fig. 6(a), we see that there is no
interference pattern even there is an overlapping between two condensates.
For the case of $g_{l}=1$ shown in Fig. 6(b), we see that low-contrast
interference patterns begin to emerge due to the interaction-induced
coherence process. In Fig. 6(c) and Fig. 6(d), we see that there are
high-contrast interference patterns. In particular, in Fig. 6(d) for $%
g_{l}=20$, two initially independent condensates can be regarded to be fully
coherent near $\protect\tau =1$.}
\end{figure}

\section{the calculations based on the second-quantization method}

The essential reason for the emergence of the interference term of two
initially independent condensates lies in that because of the exchange
symmetry of identical bosons and interatomic interaction, the two initially
independent condensates become coherent after the overlapping between the
two condensates. The physical mechanism of this interaction-induced
coherence can be understood further based on the second quantization method.
Thus, in this section, we give the density expectation value and evolution
equations based on the second-quantization method. The merit of the
second-quantization method lies in that the indistinguishability of
identical bosons is satisfied when the correct commutation relation of the
field operator is used.

\subsection{the density expectation value}

\bigskip For two initially independent condensates comprising particle
number $N_{1}$ and $N_{2}$, the corresponding quantum state is (See for
example Refs. \cite{Leggett,Pethick}):%
\begin{equation}
\left\vert N_{1},N_{2}\right\rangle =\frac{\Xi _{n}}{\sqrt{N_{1}!N_{2}!}}(%
\widehat{a}_{1}^{\dag })^{N_{1}}(\widehat{a}_{2}^{\dag })^{N_{2}}\left\vert
0\right\rangle ,  \label{initial-state}
\end{equation}%
where $\Xi _{n}$ is a normalization constant to assure $\left\langle
N_{1},N_{2}|N_{1},N_{2}\right\rangle =1$. $\widehat{a}_{1}^{\dag }$ ($%
\widehat{a}_{2}^{\dag }$) is a creation operator which creates a particle
described by the single-particle state $\phi _{1}$ ($\phi _{2}$) in the left
(right) condensate. Similarly to Eq. (\ref{wave-function}), the quantum
depletion is omitted in this sort of quantum state.

For two initially independent condensates, for the state given by Eq. (\ref%
{initial-state}), it seems that there is no interference term in the density
expectation value with the following simple calculation:
\begin{eqnarray}
n_{d}\left( \mathbf{r},t\right) &=&\left\langle N_{1},N_{2},t\right\vert
\widehat{\Psi }^{\dag }\widehat{\Psi }\left\vert N_{1},N_{2},t\right\rangle
\nonumber \\
&=&\left\langle N_{1},N_{2},t\right\vert \left( \widehat{a}_{1}^{\dag }%
\widehat{a}_{1}\left\vert \phi _{1}\left( \mathbf{r},t\right) \right\vert
^{2}+\widehat{a}_{2}^{\dag }\widehat{a}_{2}\left\vert \phi _{2}\left(
\mathbf{r},t\right) \right\vert ^{2}\right) \left\vert
N_{1},N_{2},t\right\rangle  \nonumber \\
&&+2\times \mathrm{Re}\left( \left\langle N_{1},N_{2},t\right\vert \widehat{a%
}_{1}^{\dag }\widehat{a}_{2}\left\vert N_{1},N_{2},t\right\rangle \phi
_{1}^{\ast }\left( \mathbf{r},t\right) \phi _{2}\left( \mathbf{r},t\right)
\right)  \label{nd-center} \\
&=&N_{1}\left\vert \phi _{1}\left( \mathbf{r},t\right) \right\vert
^{2}+N_{2}\left\vert \phi _{2}\left( \mathbf{r},t\right) \right\vert ^{2}.
\label{nd-field(2)}
\end{eqnarray}%
In the above equation, the field operator $\widehat{\Psi }$ is expanded as $%
\widehat{\Psi }=\widehat{a}_{1}\phi _{1}+\widehat{a}_{2}\phi _{2}+\cdots $
with $\widehat{a}_{1}$ and $\widehat{a}_{2}$ being the annihilation
operators. One should note that, to get (\ref{nd-field(2)}) from (\ref%
{nd-center}), there is an implicit assumption that $\widehat{a}_{1}$ and $%
\widehat{a}_{2}^{\dag }$ are commutative. This holds when $\int \phi
_{1}\left( \mathbf{r},t\right) \phi _{2}^{\ast }\left( \mathbf{r},t\right)
dV=0$. When $\left[ \widehat{a}_{1},\widehat{a}_{2}^{\dag }\right] =0$, it
is easy to understand that the interference term (the last term in Eq. (\ref%
{nd-center})) is zero in $n_{d}\left( \mathbf{r},t\right) $.

As shown in the previous sections, $\phi _{1}$ and $\phi _{2}$ will become
non-orthogonal in the presence of the interatomic interaction. The operators
$\widehat{a}_{1}$ and $\widehat{a}_{2}$ can be written as%
\begin{equation}
\widehat{a}_{1}=\int \widehat{\Psi }\phi _{1}^{\ast }dV,  \label{a1}
\end{equation}%
and%
\begin{equation}
\widehat{a}_{2}=\int \widehat{\Psi }\phi _{2}^{\ast }dV.  \label{a2}
\end{equation}%
Here $\widehat{\Psi }$ is the field operator. By using the commutation
relations of the field operators $[\widehat{\Psi }\left( \mathbf{r}%
_{1},t\right) ,\widehat{\Psi }\left( \mathbf{r}_{2},t\right) ]=0$ and $[%
\widehat{\Psi }\left( \mathbf{r}_{1},t\right) ,\widehat{\Psi }^{\dagger
}\left( \mathbf{r}_{2},t\right) ]=\delta \left( \mathbf{r}_{1}-\mathbf{r}%
_{2}\right) $, it is easy to get the commutation relation%
\begin{equation}
\lbrack \widehat{a}_{1},\widehat{a}_{2}^{\dagger }]=\zeta ^{\ast }.
\label{commre}
\end{equation}%
We see that $\widehat{a}_{1}$ and $\widehat{a}_{2}^{\dagger }$ are not
commutative any more for $\int \phi _{1}\phi _{2}^{\ast }dV$ being a nonzero
value. In this situation, it is obvious that one can not get the result (\ref%
{nd-field(2)}) from (\ref{nd-center}) any more. This means that one should
be very careful to get the correct density expectation value for two
initially independent condensates.

It is well-known that the field operator should be expanded in terms of a
complete and orthogonal basis set. Generally speaking, the field operator $%
\widehat{\Psi }$ can be expanded as:
\begin{equation}
\widehat{\Psi }=\widehat{a}_{1}\phi _{1}+\widehat{k}\phi _{2}^{\prime
}+\cdots ,  \label{new-expansion}
\end{equation}%
where $\phi _{1}$ and $\phi _{2}^{\prime }$ are orthogonal normalization
wave functions. Assuming that $\phi _{2}^{\prime }=\beta \left( \phi
_{2}+\alpha \phi _{1}\right) $, based on the conditions $\int \phi
_{1}^{\ast }\phi _{2}^{\prime }dV=0$ and $\int \left\vert \phi _{2}^{\prime
}\right\vert ^{2}dV=1$, we have $\left\vert \beta \right\vert =\left(
1-\left\vert \zeta \right\vert ^{2}\right) ^{-1/2}$ and $\alpha =-$ $\zeta
^{\ast }$. Based on%
\begin{equation}
\widehat{k}=\int \widehat{\Psi }\left( \phi _{2}^{\prime }\right) ^{\ast }dV,
\label{k-operator}
\end{equation}%
we have%
\begin{equation}
\widehat{a}_{2}=\frac{\widehat{k}}{\beta ^{\ast }}+\zeta \widehat{a}_{1}.
\label{a2-k}
\end{equation}%
It is easy to get the following commutation relations:
\begin{eqnarray}
\lbrack \widehat{k},\widehat{k}] &=&[\widehat{k}^{\dagger },\widehat{k}%
^{\dagger }]=0,[\widehat{k},\widehat{k}^{\dagger }]=1,  \nonumber \\
\left[ \widehat{a}_{1},\widehat{a}_{1}\right] &=&[\widehat{a}_{1}^{\dagger },%
\widehat{a}_{1}^{\dagger }]=0,[\widehat{a}_{1},\widehat{a}_{1}^{\dagger }]=1,
\nonumber \\
\lbrack \widehat{k},\widehat{a}_{1}] &=&[\widehat{k},\widehat{a}%
_{1}^{\dagger }]=0.  \label{kcom}
\end{eqnarray}

Because $\widehat{k}$ and $\widehat{a}_{1}^{\dagger }$ are commutative, it
is convenient to calculate the density expectation value $n_{d}\left(
\mathbf{r},t\right) $ by using the operators $\widehat{k}$ and $\widehat{a}%
_{1}^{\dagger }$. After straightforward derivations, the exact expression of
the density expectation value is
\begin{eqnarray}
n_{d}\left( \mathbf{r},t\right) &=&\frac{\Xi _{n}^{2}}{N_{1}!N_{2}!}%
\left\langle 0\right\vert \left[ \frac{\widehat{k}}{\beta ^{\ast }}+\zeta
\widehat{a}_{1}\right] ^{N_{2}}\widehat{a}_{1}^{N_{1}}\left[ \widehat{a}%
_{1}^{\dag }\phi _{1}^{\ast }+\widehat{k}^{\dag }\left( \phi _{2}^{\prime
}\right) ^{\ast }\right]  \nonumber \\
&&\left[ \widehat{a}_{1}\phi _{1}+\widehat{k}\phi _{2}^{\prime }\right]
\left( \widehat{a}_{1}^{\dagger }\right) ^{N_{1}}\left[ \frac{\widehat{k}%
^{\dagger }}{\beta }+\zeta ^{\ast }\widehat{a}_{1}^{\dagger }\right]
^{N_{2}}\left\vert 0\right\rangle  \nonumber \\
&=&\alpha _{d}\left\vert \phi _{1}\left( \mathbf{r},t\right) \right\vert
^{2}+2\beta _{d}\times \mathrm{Re}\left( e^{i\varphi _{c}}\phi _{1}^{\ast
}\left( \mathbf{r},t\right) \phi _{2}\left( \mathbf{r},t\right) \right)
+\gamma _{d}\left\vert \phi _{2}\left( \mathbf{r},t\right) \right\vert ^{2},
\label{ndensity}
\end{eqnarray}%
where the coefficients are
\begin{eqnarray}
\alpha _{d} &=&\sum\limits_{i=0}^{N_{2}}\frac{\Xi _{n}^{2}N_{2}!\left(
N_{1}+i-1\right) !N_{1}\left( 1-\left\vert \zeta \right\vert ^{2}\right)
^{N_{2}-i}\left\vert \zeta \right\vert ^{2i}}{i!i!\left( N_{1}-1\right)
!\left( N_{2}-i\right) !},  \label{adS} \\
\beta _{d} &=&\sum\limits_{i=0}^{N_{2}-1}\frac{\Xi _{n}^{2}N_{2}!\left(
N_{1}+i\right) !\left( 1-\left\vert \zeta \right\vert ^{2}\right)
^{N_{2}-i-1}\left\vert \zeta \right\vert ^{2i+1}}{i!\left( i+1\right)
!\left( N_{1}-1\right) !\left( N_{2}-i-1\right) !},  \label{bdS} \\
\gamma _{d} &=&\sum\limits_{i=0}^{N_{2}-1}\frac{\Xi _{n}^{2}N_{2}!\left(
N_{1}+i\right) !\left( 1-\left\vert \zeta \right\vert ^{2}\right)
^{N_{2}-i-1}\left\vert \zeta \right\vert ^{2i}}{i!i!N_{1}!\left(
N_{2}-i-1\right) !}.  \label{cdS}
\end{eqnarray}%
In addition, the normalization constant $\Xi _{n}$ is determined by%
\begin{equation}
\Xi _{n}^{2}\left( \sum\limits_{i=0}^{N_{2}}\frac{N_{2}!\left(
N_{1}+i\right) !\left( 1-\left\vert \zeta \right\vert ^{2}\right)
^{N_{2}-i}\left\vert \zeta \right\vert ^{2i}}{i!i!N_{1}!\left(
N_{2}-i\right) !}\right) =1.  \label{normconstS}
\end{equation}

The above density expectation value is obtained based on the second
quantization method. Although it seems that the expressions of the
coefficients given by Eqs. (\ref{adS}), (\ref{bdS}), (\ref{cdS}) and (\ref%
{normconstS}) are quite different from the results calculated from the
many-body wave function, we have proven that $\alpha _{d}=a_{d}$, $\beta
_{d}=b_{d}$, $\gamma _{d}=c_{d}$ and $\Xi _{n}=A_{n}$. Thus, the density
expectation value given by Eq. (\ref{ndensity}) is the same as the result
calculated from the many-body wave function $\Psi _{N_{1}N_{2}}$ which
satisfies the exchange symmetry of identical bosons.

\subsection{the evolution equations}

Based on the second quantization method, after removing the double-well
potential, the Hamiltonian in the second-quantization method is%
\begin{equation}
\widehat{H}_{s}=\int dV\left( \frac{\hbar ^{2}}{2m}\nabla \widehat{\Psi }%
^{\dag }\cdot \nabla \widehat{\Psi }+\frac{g}{2}\widehat{\Psi }^{\dag }%
\widehat{\Psi }^{\dag }\widehat{\Psi }\widehat{\Psi }\right) .
\end{equation}%
After straightforward calculations, the overall energy of the whole system
is
\begin{equation}
E=E_{kin}+E_{int},
\end{equation}%
where the kinetic energy $E_{kin}$ is given by\bigskip
\begin{eqnarray}
E_{kin} &=&\int \left\langle N_{1},N_{2},t\right\vert \frac{\hbar ^{2}}{2m}%
\nabla \widehat{\Psi }^{\dag }\cdot \nabla \widehat{\Psi }\left\vert
N_{1},N_{2},t\right\rangle dV  \nonumber \\
&=&\int dV\left( \frac{\alpha _{d}\hslash ^{2}}{2m}\nabla \phi _{1}^{\ast
}\cdot \nabla \phi _{1}+\frac{\beta _{d}\hbar ^{2}}{2m}e^{i\varphi
_{c}}\nabla \phi _{1}^{\ast }\cdot \nabla \phi _{2}\right.  \nonumber \\
&&\left. +\frac{\beta _{d}\hbar ^{2}}{2m}e^{-i\varphi _{c}}\nabla \phi
_{2}^{\ast }\cdot \nabla \phi _{1}+\frac{\gamma _{d}\hbar ^{2}}{2m}\nabla
\phi _{2}^{\ast }\cdot \nabla \phi _{2}\right) .
\end{eqnarray}%
In addition, the interaction energy $E_{int}$ of the whole system is given by%
\begin{eqnarray}
E_{int} &=&\int \left\langle N_{1},N_{2},t\right\vert \frac{g}{2}\widehat{%
\Psi }^{\dag }\widehat{\Psi }^{\dag }\widehat{\Psi }\widehat{\Psi }%
\left\vert N_{1},N_{2},t\right\rangle dV  \nonumber \\
&=&\frac{g}{2}\int dV\left[ \alpha _{1}\left\vert \phi _{1}\right\vert
^{4}+\alpha _{2}\left\vert \phi _{2}\right\vert ^{4}+\alpha _{3}\left\vert
\phi _{1}\right\vert ^{2}\left\vert \phi _{2}\right\vert ^{2}\right.
\nonumber \\
&&\left. +\mathrm{Re}\left( \alpha _{4}\left\vert \phi _{1}\right\vert
^{2}\phi _{1}^{\ast }\phi _{2}e^{i\varphi _{c}}+\alpha _{5}\left( \phi
_{1}^{\ast }\right) ^{2}\phi _{2}^{2}e^{2i\varphi _{c}}+\alpha
_{6}\left\vert \phi _{2}\right\vert ^{2}\phi _{1}^{\ast }\phi
_{2}e^{i\varphi _{c}}\right) \right] ,
\end{eqnarray}%
where the coefficients are given by%
\begin{eqnarray}
\alpha _{1} &=&\sum\limits_{i=0}^{N_{2}}\frac{\Xi _{n}^{2}N_{2}!\left(
N_{1}+i-2\right) !N_{1}\left( N_{1}-1\right) }{i!i!\left( N_{1}-2\right)
!\left( N_{2}-i\right) !}\left( 1-\left\vert \zeta \right\vert ^{2}\right)
^{N_{2}-i}\left\vert \zeta \right\vert ^{2i},  \nonumber \\
\alpha _{2} &=&\sum\limits_{i=0}^{N_{2}-2}\frac{\Xi _{n}^{2}N_{2}!\left(
N_{1}+i\right) !}{i!i!N_{1}!\left( N_{2}-i-2\right) !}\left( 1-\left\vert
\zeta \right\vert ^{2}\right) ^{N_{2}-i-2}\left\vert \zeta \right\vert ^{2i},
\nonumber \\
\alpha _{3} &=&\sum\limits_{i=0}^{N_{2}-1}\frac{4\Xi _{n}^{2}N_{2}!\left(
N_{1}+i-1\right) !N_{1}}{i!i!\left( N_{1}-1\right) !\left( N_{2}-i-1\right) !%
}\left( 1-\left\vert \zeta \right\vert ^{2}\right) ^{N_{2}-i-1}\left\vert
\zeta \right\vert ^{2i},  \nonumber \\
\alpha _{4} &=&\sum\limits_{i=0}^{N_{2}-1}\frac{4\Xi _{n}^{2}N_{2}!\left(
N_{1}+i-1\right) !N_{1}}{i!\left( i+1\right) !\left( N_{1}-2\right) !\left(
N_{2}-i-1\right) !}\left( 1-\left\vert \zeta \right\vert ^{2}\right)
^{N_{2}-i-1}\left\vert \zeta \right\vert ^{2i+1},  \nonumber \\
\alpha _{5} &=&\sum\limits_{i=0}^{N_{2}-2}\frac{2\Xi _{n}^{2}N_{2}!\left(
N_{1}+i\right) !}{i!\left( i+2\right) !\left( N_{1}-2\right) !\left(
N_{2}-i-2\right) !}\left( 1-\left\vert \zeta \right\vert ^{2}\right)
^{N_{2}-i-2}\left\vert \zeta \right\vert ^{2i+2},  \nonumber \\
\alpha _{6} &=&\sum\limits_{i=0}^{N_{2}-2}\frac{4\Xi _{n}^{2}N_{2}!\left(
N_{1}+i\right) !}{i!\left( i+1\right) !\left( N_{1}-1\right) !\left(
N_{2}-i-2\right) !}\left( 1-\left\vert \zeta \right\vert ^{2}\right)
^{N_{2}-i-2}\left\vert \zeta \right\vert ^{2i+1}.
\end{eqnarray}%
We have proven that the above overall energy is equal to the results based
on the many-body wave function, by checking that $h_{i}=\alpha _{i}$ for $%
i=1 $, $2$, $\cdots $, $6$.

When the non-orthogonal property between $\phi _{1}$ and $\phi _{2}$ are
considered, because the derivations of the density expectation value and
overall energy are quite cumbersome, the same results based on the
first-quantization method and second-quantization method give us strong
evidence that our derivations are correct.

\section{quantum fluctuations and orthogonality of the whole quantum state}

In both the first-quantization and second-quantization methods of the
previous calculations, the quantum depletion originating from the elementary
excitations at zero temperature is omitted in the quantum state of the whole
system. Based on the Bogoliubov theory of the elementary excitations, for
the initial quantum state, the number of particles due to the quantum
depletion is of the order of $\left( a/\overline{l}\right) ^{3/2}$ and thus
the quantum depletion is negligible for Bose condensate in dilute gases
considered in the present work. With the development of time, the role of
quantum depletion can also be omitted. Because of the factor $e^{\pm i%
\mathbf{k}\cdot \mathbf{r}}$ ($\left\vert \mathbf{k}\right\vert $ is the
wave number of the elementary excitations) in the wave function $\phi _{%
\mathbf{k}}$ of the elementary excitations, a simple analysis shows
qualitatively that $\left\langle \phi _{1}|\phi
_{\mathbf{k}}\right\rangle $ and $\left\langle \phi _{2}|\phi
_{\mathbf{k}}\right\rangle $ are of the order of $\left\vert \zeta
\right\vert e^{-(\left\vert \mathbf{k}\right\vert L)^{2}}$ with $L$
being the spatial size of the system. This exponential decay of
$\left\langle \phi _{1}|\phi _{\mathbf{k}}\right\rangle $ and
$\left\langle \phi _{2}|\phi _{\mathbf{k}}\right\rangle $ originates
from the integral where there is spatially oscillating phase factor
in the wave functions of the elementary excitations and condensates.
Thus, the contribution to the effective order parameter and density
expectation value due to elementary excitations can be omitted
safely.

In the preceding paragraph, it is shown that in calculating the evolution of
the density expectation value, the role of quantum depletion can be omitted.
However, the quantum depletion plays important role in the consistency of
our theory. Here we discuss mainly the orthogonality for the quantum state
of the whole system, especially about the subtle problem that whether the
nonorthogonality between $\phi _{1}$ and $\phi _{2}$ violates the
orthogonality of the whole system which must be satisfied.

This sort of problem about orthogonality exists also for a single
condensate. For a single condensate, if the quantum depletion is omitted,
the many-body wave function is%
\begin{equation}
\Psi \left( \mathbf{r}_{1},\cdots ,\mathbf{r}_{N},t\right) =\phi \left(
\mathbf{r}_{1},t\right) \cdots \phi \left( \mathbf{r}_{N},t\right) .
\label{many-body-single}
\end{equation}%
Based on the many-body Schr\H{o}dinger equation and the action principle, it
is easy to get the following GP equation%
\begin{equation}
i\hbar \frac{\partial \phi }{\partial t}=-\frac{\hbar ^{2}}{2m}\nabla
^{2}\phi +V_{ext}\left( \mathbf{r},t\right) \phi +g\left( N-1\right)
\left\vert \phi \right\vert ^{2}\phi .  \label{single-GP}
\end{equation}%
Assume that there are two orthogonally initial quantum states $\Psi
_{A}\left( t=0\right) =\phi _{A}\left( \mathbf{r}_{1},t=0\right) \cdots \phi
_{A}\left( \mathbf{r}_{N},t=0\right) $ and $\Psi _{B}\left( t=0\right) =\phi
_{B}\left( \mathbf{r}_{1},t=0\right) \cdots \phi _{B}\left( \mathbf{r}%
_{N},t=0\right) $. With the development of time, because of the
nonlinearity of the GP equation, $\phi _{A}$ and $\phi _{B}$ may
become non-orthogonal in the presence of interatomic interaction. It
is obvious that $\int \Psi _{A}^{\ast }\Psi
_{B}d\mathbf{r}_{1}\cdots d\mathbf{r}_{N}=\left[ \int \phi
_{A}^{\ast }\left( \mathbf{r},t\right) \phi _{B}\left( \mathbf{r},t\right) d%
\mathbf{r}\right] ^{N}$. Although for large $N$, this integral can be
approximated as zero, based on the consideration of consistency, the
physical mechanism of the exact orthogonality between $\Psi _{A}$ and $\Psi
_{B}$ is an interesting problem.

We think that the effect of quantum depletion omitted in Eq. (\ref%
{many-body-single}) is the essential reason for this sort of unphysical
nonorthogonality. For the Hamiltonian (\ref{hamiltonian-first}) of the whole
system, the term $g\sum_{i<j}^{N}\delta \left( \mathbf{r}_{i}-\mathbf{r}%
_{j}\right) $ is non-factorable about the coordinate, and thus the exact
solution of the many-body wave function is non-factorable too. For a single
condensate, the exact many-body wave function can be assumed as%
\begin{equation}
\Psi _{exact}\left( \mathbf{r}_{1},\cdots ,\mathbf{r}_{N},t\right) =c\phi
\left( \mathbf{r}_{1},t\right) \cdots \phi \left( \mathbf{r}_{N},t\right)
F_{N}\left( \mathbf{r}_{1},\cdots ,\mathbf{r}_{N},t\right) .
\end{equation}%
Here $F_{N}$ accounts for the non-factorable component, and $c$ is a
normalization constant. In the hard-sphere approximation, $F_{N}=0$ if $%
\left\vert \mathbf{r}_{i}-\mathbf{r}_{j}\right\vert \leq 2r$ for any $i\neq
j $ ($r$ is the hard-sphere radius), and $F_{N}=1$ for otherwise situation. $%
F_{N}$ represents the quantum depletion, and was successfully used
to calculate the quantum depletion of superfluid liquid
$^{4}\mathrm{He}$ in Ref. \cite{Penrose}. The omission of the
quantum depletion means that we approximate $F_{N}$ as $1$. For
dilute Bose condensed gases, $r<<\overline{l} $. Thus, the wave
function (\ref{many-body-single}) can describe well the condensate.
When the orthogonality of the whole system is considered, we stress
here that the wave function $\Psi _{exact}$ should be used. The
non-factorable factor $F_{N}$ assures the orthogonality of the
many-body quantum state, because $\Psi _{exact}$ is the exact
solution of the many-body Schr\H{o}dinger equation. For dilute Bose
condensed gases, these analyses lead to two results: when the
dynamic evolution of dilute Bose condensed gases is considered,
omitting the quantum depletion can give us quite good description;
when the consistency especially the orthogonality is considered, one
should consider the role of quantum depletion.

It is natural to generalize the above analyses to two initially independent
condensates. The nonorthogonality between $\phi _{1}$ and $\phi _{2}$ does
not mean in any sense the violation of the orthogonality of the quantum
state of the whole system. We stress here again that when orthogonality is
considered, we should check whether the quantum state of the whole system
satisfies the orthogonality. When the quantum depletion is considered, the
many-body wave function takes the following form%
\begin{equation}
\Psi _{N_{1}N_{2}}^{e}\left( \mathbf{r}_{1},\mathbf{r}_{2},\cdots ,\mathbf{r}%
_{N},t\right) =c\Psi _{N_{1}N_{2}}\left( \mathbf{r}_{1},\mathbf{r}%
_{2},\cdots ,\mathbf{r}_{N},t\right) F_{N}\left( \mathbf{r}_{1},\cdots ,%
\mathbf{r}_{N},t\right) .
\end{equation}%
Here $\Psi _{N_{1}N_{2}}$ is given by Eq. (\ref{wave-function}).
$\Psi _{N_{1}N_{2}}^{e}$ satisfies exactly the many-body
Schr\H{o}dinger equation, and thus this sort of wave function
satisfies the orthogonality condition. Assume that $\Psi
_{N_{1}N_{2}}^{e-A}$ is constructed by $\phi _{1}^{A}$ and $\phi
_{2}^{A}$, while $\Psi _{N_{1}N_{2}}^{e-B}$ is constructed by $\phi
_{1}^{B}$ and $\phi _{2}^{B}$. Assume further that $\phi _{1}^{A}$,
$\phi _{2}^{A}$, $\phi _{1}^{B}$ and $\phi _{2}^{B}$ are orthogonal
to each other at the initial time, so that $\Psi
_{N_{1}N_{2}}^{e-A}$ and $\Psi _{N_{1}N_{2}}^{e-B}$ are orthogonal
at the initial time. Omitting the quantum depletion and roughly
speaking, the integral between $\Psi _{N_{1}N_{2}}^{e-A}$ and $\Psi
_{N_{1}N_{2}}^{e-B}$ is of the order of $\left[ \int \left( \phi
_{1}^{A}\right) ^{\ast }\phi _{1}^{B}d\mathbf{r}\right] ^{\alpha
}\left[ \int \left( \phi _{1}^{A}\right) ^{\ast }\phi
_{2}^{B}d\mathbf{r}\right]
^{\beta }\left[ \int \left( \phi _{2}^{A}\right) ^{\ast }\phi _{1}^{B}d%
\mathbf{r}\right] ^{\gamma }\left[ \int \left( \phi _{2}^{A}\right) ^{\ast
}\phi _{2}^{B}d\mathbf{r}\right] ^{\delta }$ (here $\alpha +\beta +\gamma
+\delta =N$). We see that similarly to a single condensate, the
orthogonality between $\Psi _{N_{1}N_{2}}^{e-A}$ and $\Psi
_{N_{1}N_{2}}^{e-B}$ is quite good for large $N$ when the quantum depletion
is omitted. But to assure the exact orthogonality of the whole quantum
state, we must consider the role of quantum depletion. Although we do not
provide here a fully rigorous proof, these analyses lead to the result that:
although for dilute Bose condensed gases, the omission of the quantum
depletion will not play important contribution to the evolution of $\phi
_{1} $ and $\phi _{2}$, it's the quantum depletion that assure the
orthogonality of the whole quantum state.

Up to date, we mainly use the bases $\phi _{1}$ and $\phi _{2}$ to study the
dynamic evolution and density expectation value. Considering the
nonorthogonality between $\phi _{1}$ and $\phi _{2}$ in the presence of
interparticle interaction, it is attractive to discuss the physical picture
using the orthogonal bases $\phi _{1}$ and $\phi _{2}^{\prime }$ introduced
in Sec. VII. If we use the orthogonal bases $\phi _{1}$ and $\phi
_{2}^{\prime }$, the quantum state of the whole system is $\left\vert
N_{1},N_{2}\right\rangle \sim (\widehat{a}_{1}^{\dag })^{N_{1}}(\widehat{k}%
^{\dagger }/\beta +\zeta ^{\ast }\widehat{a}_{1}^{\dagger
})^{N_{2}}\left\vert 0\right\rangle \sim \sum_{m=0}^{N_{2}}C_{m}(\widehat{a}%
_{1}^{\dag })^{N_{1}+m}(\widehat{k}^{\dag })^{N_{2}-m}\left\vert
0\right\rangle $. We see that the number of particles in the orthogonal
modes $\phi _{1}$ and $\phi _{2}^{\prime }$ are no more definite. This
quantum state becomes a superposition of different number of particles in
the orthogonal modes $\phi _{1}$ and $\phi _{2}^{\prime }$. This result is
very natural because the coupling (interaction) between two initially
independent condensates leads to the coherent transfer of particles between
two condensates. In Ref. \cite{Barnett}, it was shown that for a single
condensate, a coherent state description of the Bose condensed system is a
robust state in the presence of the interactions between the condensate and
its environment. In the present work, our research shows that the
interaction between two condensates makes each condensate become a coherent
superposition of different particle number. The results in Ref. \cite%
{Barnett} about the robustness of the coherent state imply that $%
\sum_{m=0}^{N_{2}}C_{m}(\widehat{a}_{1}^{\dag })^{N_{1}+m}(\widehat{k}^{\dag
})^{N_{2}-m}\left\vert 0\right\rangle $ is a robust quantum state even when
the coupling with the environment exists.

\section{summary and discussion}

In summary, upon expansion, we calculate the density expectation value of
two initially independent condensates. It is found that there is a nonzero
interference term when the interatomic interaction and the exchange symmetry
of identical bosons are both considered carefully. In fact, it is well-known
that the interaction plays an essential role in the formation of the order
parameter of Bose-condensed gases, i.e. the formation of a stable coherent
property. Here, we provide an example in which the interaction induces
coherent evolution between two initially independent condensates.

In the present popular viewpoint, the high-order correlation function and
quantum measurement process are used to interpret the interference patterns
observed for two initially independent condensates. Although we find that
there is already high-contrast interference patterns in the density
expectation value for the experimental parameters in Ref. \cite{Andrew}, it
is still possible that the high-order correlation function and thus the
mechanism of the interference due to the measurement process play an
important role in the emergence of the interference patterns. Thus, to
investigate more clearly the interference pattern due to the measurement
process alone, we believe an experimental investigation of two independent
ideal condensates would be very interesting, because there is no
interference term for ideal condensates in the density expectation value. In
the last few years, the rapid experimental advances of Feshbach resonance
where the scattering length can be tuned from positive to negative make this
sort of experiment be feasible.

Note added: The present paper is an expansion of our two previous
unpublished works \cite{xiong1,xiong2}. In addition, after the
submission of the present paper, we noticed a paper \cite{ceder} by
L. S. Cederbaum et al., where the role of interaction in the
emergence of interference patterns for two initially independent
Bose condensates is stressed.

\begin{acknowledgments}
This work is supported by NSFC under Grant Nos. 10474117, 10474119 and NBRPC
under Grant Nos. 2005CB724508, 2001CB309309.
\end{acknowledgments}


\begin{thebibliography}{99}
\bibitem{Nature} See special issue Nature Insight: 2002 \textit{Ultracold
Matter, Nature} (London) \textbf{416} 205

\bibitem{Anderson} Anderson M H, Ensher J R, Mathews M R, Wieman C E and
Cornell E A 1995 \textit{Science} \textbf{269} 198

\bibitem{Davis} Davis K B, Mewes M-O, Andrews M R, van Druten N J, Durfee D
S, Kurn D M and Ketterle W 1995 \textit{Phys. Rev. Lett.} \textbf{75} 3969

\bibitem{Bradley} Bradley C C, Sackett C A, Tollett J J and Hulet R G 1995
\textit{Phys. Rev. Lett.} \textbf{75} 1687

\bibitem{Andrew} Andrews M R, Townsend C G, Miesner H J, Durfee D S, Kurn D
M and Ketterle W 1997\textit{\ Science} \textbf{275} 637

\bibitem{ketterle} Ketterle W, Durfee D S and Stamper-Kurn D M 1999 in
\textit{Making, probing and understanding Bose-Einstein condensates, }%
Proceedings of the International School of Physics, \textquotedblleft Enrico
Fermi\textquotedblright\ Course CXL, edited by Inguscio M, Stringari S and
Wieman C E (IOS Press: Amsterdam), pp. 67--359

\bibitem{Leggett} Leggett A J 2001 \textit{Rev. Mod. Phys.} \textbf{73} 307

\bibitem{Pethick} Pethick C J and Smith H 2002 \textit{Bose-Einstein
Condensation in Dilute Gases} (Cambridge University: Cambridge)

\bibitem{Stringari} Pitaevskii L P and Stringari S 2003 \textit{%
Bose-Einstein condensation} (Clarendon: Oxford)

\bibitem{JAV} Javanainen J and Yoo S M 1996 \textit{Phys. Rev. Lett.}
\textbf{76} 161

\bibitem{CASTIN} Castin Y and Dalibard J 1997 \textit{Phys. Rev. A} \textbf{%
55} 4330

\bibitem{Zoller} Cirac J I, Gardiner C W, Naraschewski M and Zoller P 1996
\textit{Phys. Rev. A} \textbf{54} R3714

\bibitem{Gross} Gross E P 1961 \textit{Nuovo Cimento} \textbf{20} 454; 1963
\textit{J. Math. Phys.} \textbf{4} 195

\bibitem{Pitae} Pitaevskii L P 1961 \textit{Zh. Eksp. Teor. Fiz.} \textbf{40}
646 [1961 \textit{Sov. Phys.-JETP} \textbf{13} 451]

\bibitem{RMP} Dalfovo F, Giorgini S, Pitaevskii L P and Stringari S 1999
\textit{Rev. Mod. Phys.} \textbf{71} 463

\bibitem{Rohrl} R\H{o}hrl A, Naraschewski M, Schenzle A and Wallis H 1997
\textit{Phys. Rev. Lett.} \textbf{78} 4143

\bibitem{Liu} Liu W M, Wu B and Niu Q 2000 \textit{Phys. Rev. Lett.} \textbf{%
84} 2294

\bibitem{Quam} Laundau L D and Lifshitz E M 1997 \textit{Quantum mechanics}
(Pergamon: Oxford)

\bibitem{cigar} Mewes M O, Andrews M R, van Druten N J, Kurn D M, Durfee D S
and Ketterle W 1996 \textit{Phys. Rev. Lett.} \textbf{77} 416

\bibitem{Penrose} Penrose O and Onsager L 1956 \textit{Phys. Rev.} \textbf{%
104} 576.

\bibitem{Barnett} Barnett S M, Burnett K and Vaccaro J A 1996 \textit{J.
Res. Natl. Inst. Stand. Technol.} \textbf{101} 593

\bibitem{xiong1} Xiong H W, Liu S J and Zhan M S 2005 \textit{Preprint} cond-mat/0507354
\bibitem{xiong2} Xiong H W and Liu S J 2005 \textit{Preprint} cond-mat/0507509
\bibitem{ceder} Cederbaum L S, Streltsov A I, Band Y B and Alon O E 2006 \textit{Preprint}
cond-mat/0607556

\end{thebibliography}
\end{document}